\pdfoutput=1

\documentclass[%
 reprint,
superscriptaddress,
 amsmath,amssymb,
 aps,
]{revtex4-2}

\usepackage{graphicx}
\usepackage{dcolumn}
\usepackage{bm}

\begin{document}


\title{Reverse-time analysis and boundary classification of directional biological dynamics with multiplicative noise}

\author{Nicolas Lenner}
 \altaffiliation[Currently at ]{Simons Center for Systems Biology, School of Natural Sciences, Institute for Advanced Study, Princeton, New Jersey, USA.}
  \email{Lenner@ias.edu}
 \affiliation{Max Planck Institute for Dynamics and Self-Organization, Göttingen, Germany}

\author{Matthias Häring}
 \affiliation{Max Planck Institute for Dynamics and Self-Organization, Göttingen, Germany}
\affiliation{Göttingen Campus Institute for Dynamics of Biological Networks, University of Göttingen, Göttingen, Germany}

\author{Stephan Eule}%
\affiliation{Max Planck Institute for Dynamics and Self-Organization, Göttingen, Germany}
\affiliation{German Primate Center—Leibniz Institute for Primate Research, Goettingen, Germany}

\author{Jörg Großhans}
\affiliation{Göttingen Campus Institute for Dynamics of Biological Networks, University of Göttingen, Göttingen, Germany}
\affiliation{Department of Biology, Philipps University Marburg, Marburg, Germany}

\author{Fred Wolf}
 \email{Fred.Wolf@ds.mpg.de}
\affiliation{Max Planck Institute for Dynamics and Self-Organization, Göttingen, Germany}
\affiliation{Göttingen Campus Institute for Dynamics of Biological Networks, University of Göttingen, Göttingen, Germany}
\affiliation{Max Planck Institute for Multidisciplinary Sciences, Göttingen, Germany}
\affiliation{Institute for the Dynamics of Complex Systems, University of Göttingen, Göttingen, Germany}
\affiliation{Center for Biostructural Imaging of Neurodegeneration, Göttingen, Germany}
\affiliation{Bernstein Center for Computational Neuroscience Göttingen, Göttingen, Germany}


\begin{abstract}
The dynamics of living systems often serves the purpose of reaching functionally important target states. We previously proposed a theory to analyze stochastic biological dynamics evolving towards target states in reverse time. However, a large class of systems in biology can only be adequately described using state-dependent noise, which had not been discussed.  For example, in gene regulatory networks, biochemical signaling networks or neuronal circuits, count fluctuations are the dominant noise component. We characterize such dynamics as an ensemble of target state aligned (TSA) trajectories and characterize its temporal evolution in reverse-time by generalized Fokker-Planck and stochastic differential equations with multiplicative noise. We establish the classification of boundary conditions for target state modeling for a wide range of power law dynamics, and derive a universal low-noise approximation of the final phase of target state convergence. Our work expands the range of theoretically tractable systems in biology and enables novel experimental design strategies for systems that involve target states.
\end{abstract}

\maketitle


\section{INTRODUCTION}

The dynamics of living and cognitive systems often serves the purpose of reaching functionally important target states\cite{sha2003hysteresis,pomerening2003building,coudreuse2010driving,rata2018two,schwarz2018precise,domingo2011switches,nachman2007dissecting,pardee1974restriction,ahrends2014controlling,maamar2007noise,xiong2003positive,losick2008stochasticity,balazsi2011cellular,hanes1996neural,hanks2015distinct,ratcliff2016diffusion,ratcliff2008diffusion,brunton2013rats,hanks2015distinct,churchland2011variance,roitman2002response}. In the dynamics of decision-making, in morphogenesis, or in cell fate determination for instance, reaching the target state terminates the respective dynamical state and irreversibly commits the system to a novel action or dynamical state. The fact that such target states in a given system are reached only once poses interesting challenges for the data-driven identification and the mathematical modeling of target-state-directed dynamics.

In Lenner et al.~\cite{lenner2023reversetime} we developed a mathematical formalism to describe and infer such dynamics in reverse time, following the dynamics from the time of target state arrival into the past and measuring time as "time to completion". We derived and characterized reverse-time Fokker-Planck and stochastic differential equations for target-state-aligned (TSA) ensembles, showed that they exhibit distinct types of universal behavior during target state convergences, and showed that this formalism enables the precise identification of the dynamical laws of target state convergence from ensemble measurements.

In a wide class of biological systems, including gene regulatory networks, biochemical signaling networks or neuronal circuits for instance, count fluctuations are the dominant noise component, which in stochastic dynamics models is best described by state-dependent, "multiplicative" noise terms\cite{LIMA2018, laing2009stochastic, gillespie2000chemical, wright1931evolution, fisher1915evolution}.
To analyze and model target-state-directed dynamics in such systems, we here develop a theory for reverse-time stochastic dynamics and TSA ensembles for systems with state-dependent noise. We characterize TSA ensembles under multiplicative noise by generalized Fokker-Planck and stochastic differential equations in reverse time, establish the classification of boundary conditions for target state modeling for a wide range of power law dynamics, and derive a universal low-noise approximation of the final phase of target state convergence.

\section{MATHEMATICAL FORMULATION OF TIME REVERSAL AND TARGET STATE ALIGNMENT}

\begin{figure}
\centering
\includegraphics[width=1.0\linewidth]
{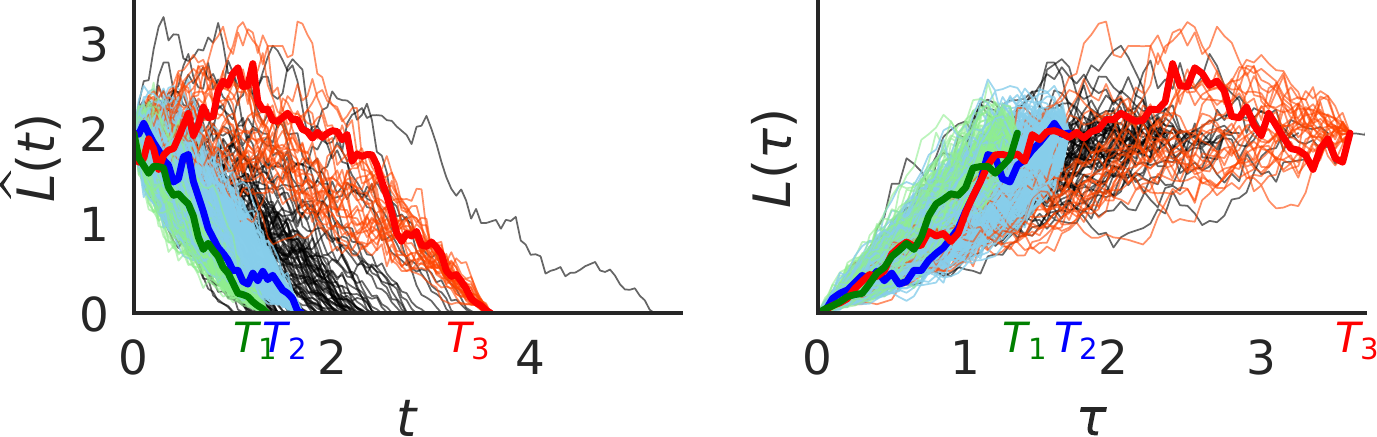}
\caption{
\textbf{
Construction of the aligned time reversed ensemble from sub-ensembles of 
different lifetimes.} 
\textbf{(Left)}: The full forward ensemble is split into sub-ensembles with different completion times $T_i$. 
We show three exemplary cases in green ($T_1$), blue ($T_2$) and red ($T_3$). To guide the eye, one sample path per sub-ensemble is highlighted. 
\textbf{(Right)}: After target state alignment and time reversal all sub-ensembles together form the new ensemble 
$P(L,\tau)$.
}
\label{fig:TSA_example}
\end{figure}

In this study, we examine the temporal evolution of a target state directed variable $\widehat{L}$ subject to both deterministic and random forcing. In the overdamped limit a Langevin equation (Ito interpretation \cite{van1981ito})
\begin{align}
\label{eq:fwd_langevin}
    d\widehat{L} = f(\widehat{L}) dt + \sqrt{D(\widehat{L})} \; dW_t
\end{align}
captures such dynamics. In our notation $\widehat{L}$ denotes a variable evolving in the usual forward time while $L$ stands for a variable that evolves in reverse time, which is introduced below. $f(\widehat{L})$ is a continuous function of $\widehat{L}$ and denotes the deterministic contribution to the force. $D(\widehat{L})$ is a position dependent diffusion "constant" and a continuous function as well. The Wiener increment implies that the noise is delta correlated with $D(\widehat{L}) \langle dW_t dW_{t'}\rangle= D(\widehat{L})\delta(t-t') dt$. The dynamics is assumed to originate from a distribution $P^{\mathrm{in}}(\widehat{L})=P(\widehat{L}_0,t_0)$. The target state is defined as $\widehat{L}_\mathrm{ts}$ and mathematically corresponds to an absorbing boundary condition. An exemplary realization of Eq.~\eqref{eq:fwd_langevin} is shown in Fig.~\ref{fig:TSA_example}.

The mathematical formulation of target state alignment and time reversal of ensembles generated by Eq.~\eqref{eq:fwd_langevin} is an intricate problem. The construction idea is as follows (see Fig.~\ref{fig:TSA_example}): 

I) In the first step we partition the forward ensemble into sub–ensembles. Each sub-ensemble $E_i$ contains only sample paths which reach the target state in the temporal interval $\left[T_i,T_i +\Delta T\right]$. The time $T_i$ is itself a stochastic variable which is distributed according to the hitting time distribution. 

II) In the second step, we revert the direction of time for each sub–ensemble. The mathematical formalism that allows the time-reversal of ensembles of identical temporal extension $T_i$ is called the reverse–time Fokker–Planck equation or equivalency the reverse–time Langevin equation \cite{anderson1982reverse}. For multiplicative noise the latter can be written as
\begin{align}
    dL =
    \bigg(
       &-f(L) \notag \\
       &+ D(L)
       \frac{\partial }{\partial L} 
       \log\left(D(L)P^{\mathrm{fw}}(L,T-\tau) \right)
    \bigg) d\tau
    \notag
    \\
    &+ \sqrt{D(L)} \; dW_{\tau}
\end{align}
in Ito interpretation. Note that $L$ is missing the hat now and we introduced the reverse time $\tau$. $P^{\mathrm{fw}}(L,T-\tau)$ is the probability density of the forward process with "fixed" forward hitting time $T$ and $\tau$ starting from zero. Hence, the initial conditions of this reverse-time dynamic are the target state $(L_\mathrm{ts},\tau_0)$ and the final state is the initial state of the forward process $(L_f, \tau_f)=(\widehat{L}_0,t_0)$. We denote the probability distribution which solves the corresponding Fokker-Planck equation as $R(L,\tau|T;L_f)$. Initial conditions at the target state 
are implied.

III) In the last step we stitch these time–reversed sub–ensembles together to form the full TSA ensemble. The alignment is achieved by choosing the same initial times $\tau_0$ for each time–reversed sub–ensemble. Each sub–ensemble is already normalized by construction. To achieve normalization in the full TSA ensemble, each sub–ensemble must be re-weighted with the hitting time distribution, the measure which counts how many trajectories are in one bin $\left[T_i,T_i +\Delta T\right]$. 
The full TSA ensemble then reads
\begin{align}
\label{sup_rtsa_delta}
 R(L,\tau) 
  &=
\notag \\
=
\int_{L_\mathrm{ts}}^\infty
&dL_f
 P^\mathrm{in}(L_f)
 \int_{\tau}^{\infty}
 dT 
 \
 R(L,\tau|T;L_f)
 \rho(T|L_f)
 \
 \ ,
\end{align}
%
%
%
where we set $\tau_0 =0$. In addition, we averaged the expression with respect to the initial value distribution $P^\mathrm{in}(L_f)$ of the forward process.

The temporal evolution of Eq.~\eqref{sup_rtsa_delta} can be obtained by taking the partial derivative of Eq.~\eqref{sup_rtsa_delta} with respect to the reverse time $\tau$. After some transformations, detailed in the supplementary information, we obtain a Fokker-Planck equation with a sink term proportional to the hitting time distribution. The corresponding Langevin equation in Ito representation reads
\begin{align}
\label{sup_tr_lv_tsa_mult}
  &dL(\tau) =
\Bigg(
      f(L) 
      +
      \notag\\
      &D(L) 
      \  
      \frac{\partial}{\partial 
L}
 \log\left(
  \int_{L_\mathrm{ts}}^{L} dL' \;
   e^{-
  \int^{L'}
 \frac{2 f(L'')}{D(L'')} dL''
 }
H(L')
 \right)  
 \Bigg)
       d\tau
\notag \\
      &+
      \sqrt{D(L)} \ d W_\tau
\end{align}
with
\begin{align}
\label{eq:H_function}
H(L')
=
 \left(
 1
 -
 \int_{L_\mathrm{ts}}^{L'} P^{\mathrm{in}}(L'') dL''
 \right) 
\end{align}
and killing measure \cite{holcman2005survival,schuss2015brownian,erban2007reactive}
\begin{align}
\label{sup_kill_tsa_fp}
 k(L,\tau)
 =\frac{
   P^\mathrm{in}(L)
 \rho_{L_\mathrm{ts}}(\tau|L)
 }{
 R_{L_\mathrm{ts}}(L,\tau;\tau_0)
 }
 \ .
\end{align}
The latter is responsible to terminate trajectories proportional to the hitting time distribution. The term $H(L)$
denotes the dependence on the initial conditions

For dynamics with target states sufficiently far from the initial conditions ($H(L) = 1)$, and temporally well separated from the start of the forward dynamics $(k(L,\tau) = 0)$, the Langevin equation simplifies to an initial value problem.

In the vicinity of target states, here chosen as $L_\mathrm{ts} = 0$, we can expand both the force law and the noise in power laws. Keeping only the terms of lowest order, the forward Langevin equation reads
\begin{align}
\label{eq:fwd_langevin_powerlaw}
    d\widehat{L} = -\gamma \widehat{L}^\alpha dt + \sqrt{D\widehat{L}^\beta} \; dW_t
 \ .
\end{align}

\section{FELLER BOUNDARY CLASSIFICATION}

Unlike for constant noise dynamics with target state directed forces ($\gamma>0 \ , \beta=0$) these dynamics are not guaranteed to always reach the target state in on average finite time. This statement can be substantiated using the Feller boundary classification scheme \cite{feller1952parabolic,ito1996diffusion,taylor1981second}.

The Feller boundary classification scheme is based on the formalization of two questions: I) What is the probability of first passage through the boundary? II) What is the average time it takes to get there from an initial state? The answer to these questions, allows to distinguish 4 scenarios: A boundary is called \textit{regular}, if it can be reached and crossed from both sides in finite time and points in the interval are accessible if starting from the boundary. \textit{Exit} boundaries can be reached in finite time, but entering the interval from the endpoint is impossible. \textit{Entrance} boundaries can not be reached in finite time from inside the defining interval but entering the interval is possible if the dynamics starts at the boundary. \textit{Natural} boundaries can neither be reached nor exited from in finite time. Feller's scheme is based on a clever mapping of arbitrary one-dimensional stochastic dynamics to the Brownian motion case. This transformation yields four criteria for which the asymptotic behavior is analytically accessible
even if it is not possible to find an explicit expression for the transition density \cite{feller1952parabolic}. For the Langevin Eq.~\eqref{tr_lv_tsa_powerlaw}, the different regimes have been calculated (see supplementary information) and are shown in Fig.~\ref{fig:boundary_approach}.

\begin{figure}
\centering
\includegraphics[width=1.0\linewidth]
{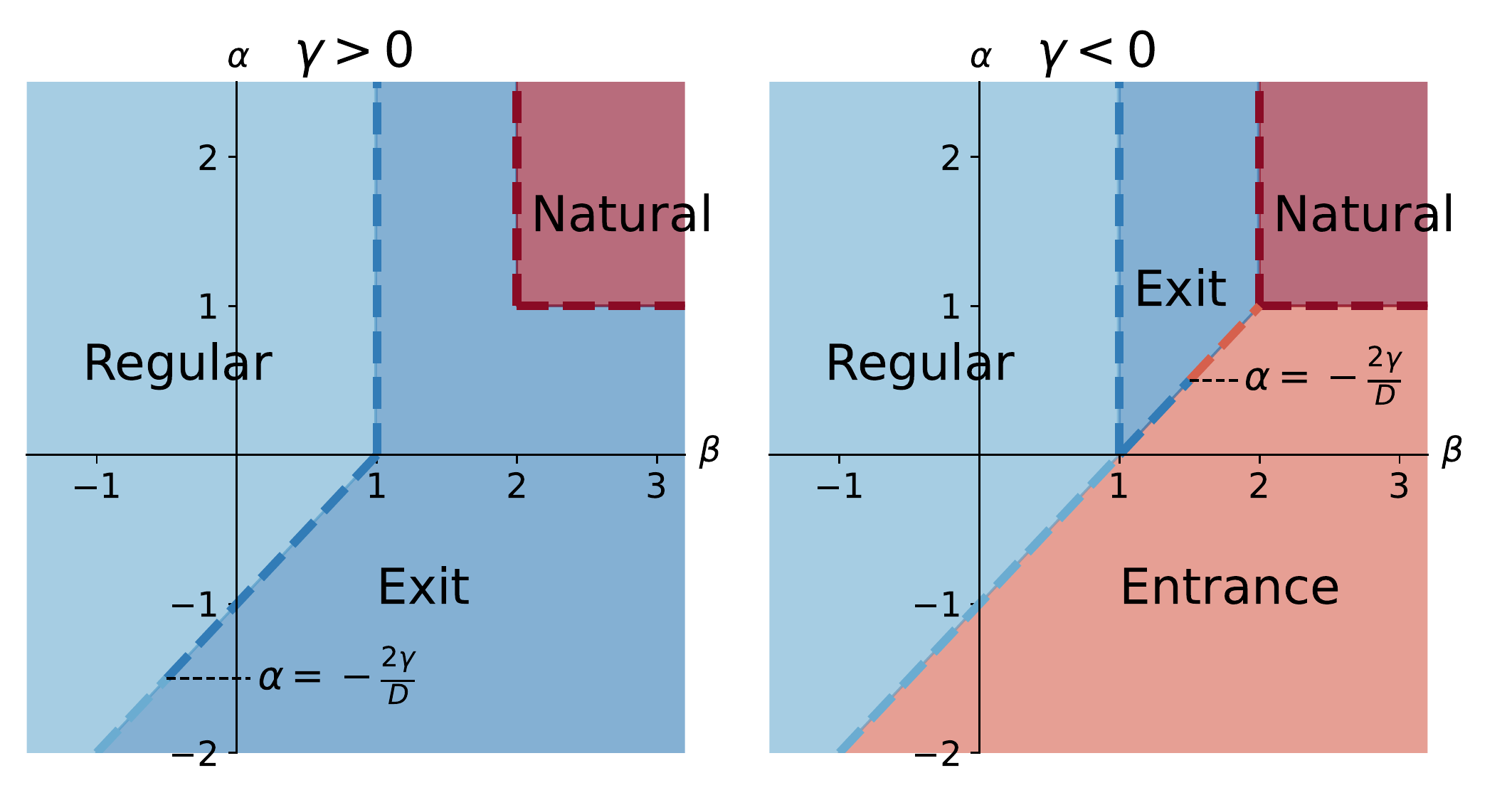}
\caption{\textbf{Target states can only be reached by a subset of dynamics with power law like deterministic and random forcing.}
Feller Boundary classification scheme \cite{feller1952parabolic, taylor1981second, trabelsi2016boundary} for the dynamics in Eq. \eqref{eq:fwd_langevin_powerlaw} and a target state at $\widehat{L}_{ts}=0$. Blue colors denote regular and exit boundaries, i.e. those cases where the boundary is reached in finite time and red colors denote natural and entrance boundaries, i.e. those cases where the target state can not be reached in finite time. Margins between the phase-regions are assigned dashed lines in the color of its respective boundary class. For a derivation of the boundary classification see the supplementary information, which also gives the definition of the different classes. \textbf{(Left):} Here we consider noise dominated target state approach ($\gamma>0$). The point at $\alpha = -\frac{2\gamma}{D}$ belongs to the exit boundary class. \textbf{(Right):} Here we consider target state directed forces ($\gamma<0$). The point at $\alpha = -\frac{2\gamma}{D}$ belongs to the entrance boundary class, the point at $\alpha=1, \beta=2$ to the natural boundary class and the point at $\alpha=0, \beta=1$ to the exit boundary class.  
}
\label{fig:boundary_approach}
\end{figure}

 To our knowledge such a general characterization as a function of exponents $\alpha$ and $\beta$ has not been given previously. Specific choices of $\alpha$ and $\beta$ have been discussed in the literature \cite{edgar2011bessel,trabelsi2016boundary, albanese2007transformations}. 
 In the supplementary information we also include an introduction to the Feller boundary classification scheme, including an intuitive explanation of the boundary classes and criteria.

Dynamics which can reach target states are either equipped with regular boundary conditions or exit boundaries. From Fig.~\ref{fig:boundary_approach} we read off, that dynamics, where the forces point towards the target state ($\gamma >0$), always reach the target state in finite time if neither the spatial derivative of the (deterministic) force nor the spatial derivative of the random force 
are zero at the boundary (that is e.g. for jointly $\alpha < 1$ and $\beta <2$). In simpler words, the boundary is not attainable if the last step towards the boundary is neither deterministically nor stochastically possible, as both terms asymptotically approach zero. 
If the force points away from the boundary, the random forcing must dominate the deterministic forcing to nevertheless reach the boundary. 
The boundary is only reached if the noise contribution is at least as large as the force contribution of the last step.

In models with the target state accessible for power law driven dynamics (Eq.~\eqref{tr_lv_tsa_powerlaw}), we determined the exact expression for the forces in the TSA reverse-time ensemble using Eq.~\eqref{sup_tr_lv_tsa_mult}. 
The reverse-time Langevin equation reads
\begin{align}
\label{tr_lv_tsa_powerlaw}
  dL(\tau) = 
\left(
      -\gamma L^\alpha 
      +
      f^\mathcal{F}(L)
 \right)
       d\tau
      +
      \sqrt{D L^\beta} \ d W_\tau
\end{align}
\begin{align}
  &\mathrm{with} \qquad
f^\mathcal{F}(L) =
\notag \\
\label{eq:freeEforce}
&= L^\beta
 \frac{D(\alpha-\beta+1) \left(-\frac{2 \gamma }{D(\alpha-\beta+1)}\right)^{\frac{1}{\alpha-\beta+1}} e^{\frac{2 \gamma  L^{\alpha-\beta+1}}{D(\alpha-\beta+1)}}}{\Theta(\alpha-\beta+1)  \Gamma \left(\frac{1}{\alpha-\beta+1}\right)-\Gamma
   \left(\frac{1}{\alpha-\beta+1},-\frac{2 L^{\alpha-\beta+1} \gamma }{D(\alpha-\beta+1)}\right)}
   \notag \\
    &\mathrm{valid \ for} \qquad \alpha -\beta \neq -1
   \ ,
\end{align}
where $\Gamma(z)$ is the gamma-function, $\Gamma \left( \nu,z \right)$ the upper incomplete gamma function and $\Theta(\nu)$ the Heaviside step function.

The case $\alpha-\beta=-1$ is substantially simpler. For this special case the reverse-time Langevin equation becomes 
\begin{align}
  \label{sup_freeEforce_fwd_powerlaw_mult_minone}
dL(\tau)
=
(\gamma+D) L^\alpha
        d\tau
      +
      \sqrt{D L^{\alpha+1}} \ d W_\tau
      \qquad \alpha -\beta = -1
     \ .
\end{align}

We solved the exact target state aligned reverse time Fokker-Planck equation for the case $\alpha = 0, \beta =1$ and present the calculated mean and variance in Fig.~\ref{fig:exact_mean_var_killScheme} (see supplementary information for details). 
\begin{figure}
\centering
\includegraphics[width=1.0\linewidth]
{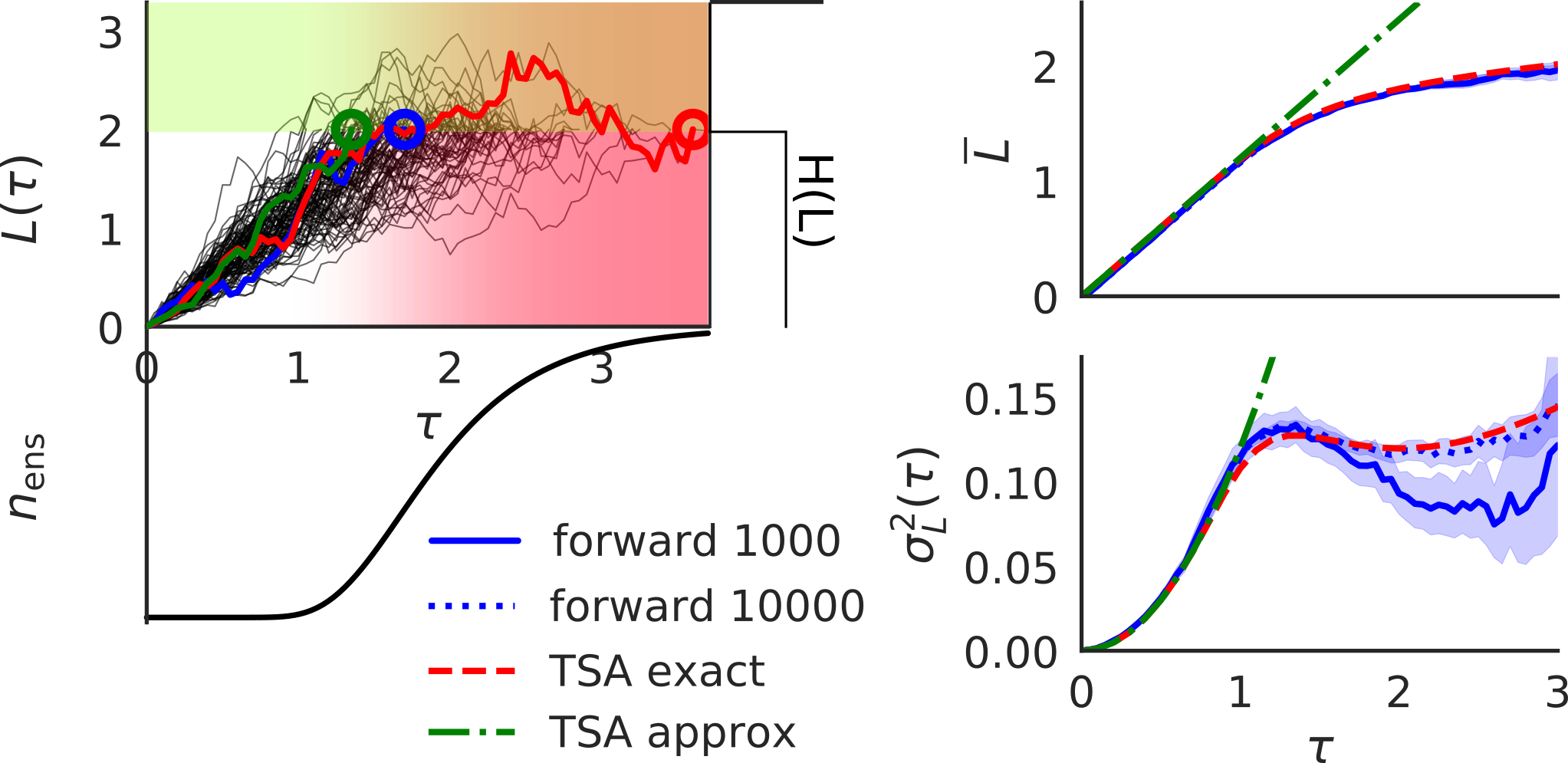}
\caption{
\textbf{The TSA reverse-time Langevin equation exactly describes the target state aligned ensemble.}
\textbf{(Left):} Schematic depiction of the dependency of the TSA dynamics Eq.~\eqref{sup_tr_lv_tsa_mult} on the forward initial condition $H(L)$ Eq.~\eqref{eq:H_function} and the killing measure $k(L,\tau)$ Eq.~\eqref{sup_kill_tsa_fp}. In the green region, i.e. above the bulk of the initial distribution measured by the sigmoidal $H(L)$, we find $f^\mathcal{F}(L)\approx 0$. Below $f^\mathcal{F}(L)$ contributes in full strength. In $\tau$ direction, an increase in the red gradient indicates, that more and more trajectories are killed.
The residual white $L-\tau$ plane close to the target state defines the region where
$k(L,\tau)\approx0$ and $H(L)=1$ holds, and the approximation of well separated initial and final states holds. Circles mark the killing of example trajectories.
\textbf{(Right):} Comparison of the forward (blue), sub-ensemble based (red), exact (black)  and approximate (green) reverse-time dynamics for $f(L)=-\gamma/L$.
Shown are the mean (Top) and variance (Bottom) of all four cases. 
95\% bootstrap confidence intervals \cite{diciccio1996bootstrap} are shown for the cases involving sampling. To exclude numerical 
inaccuracies due to rarely visited tails of the distribution of completion 
times $\rho(T|L_f)$, we directly sampled $T_i$ from the numerically 
obtained hitting time distribution of the forward process. Results 
were obtained using each 1000 sample path realizations with 
parameter settings $\gamma = 1$, $D =0.2$, $\widehat{L}_\mathrm{init}=2$.
}
\label{fig:exact_mean_var_killScheme}
\end{figure}

The TSA approximation, valid for dynamics close to target states ($H(L) = 1$, no killing measure), can be evaluated for all $\alpha = \beta -1$ ($\beta<2$). We made use of results for Bessel processes \cite{bray2000random,edgar2011bessel}. 
We calculated the mean and variance (see supplementary information) and compare to both the exact result and the mean and variance obtained by forward simulation and subsequent target state alignment and time reversal. The simulations are in excellent agreement with the theoretical expressions (see Fig.~\ref{fig:exact_mean_var_killScheme}). The TSA approximation, evaluated analytically with respect to mean and variance, captures the dynamics close to the target state with similar quality (see supplementary information for calculations and Fig.~\ref{fig:exact_mean_var_killScheme}). This demonstrates the effectiveness and power of the TSA approximation.

\section{UNIVERSAL BEHAVIOR CLOSE TO TARGET STATES}

Close to the target state the free energy force (Eq.~\eqref{eq:freeEforce}) can be expanded in low orders of $L$. We find two easily interpretable cases. The reverse-time Langevin equation for power laws (Eq.~\eqref{tr_lv_tsa_powerlaw} \& Eq.~\eqref{sup_freeEforce_fwd_powerlaw_mult_minone}) simplifies to a noise driven ($\alpha \ge \beta$)
\begin{align}
\label{sup_tr_lv_tsa_sn_multN_alpha_ge_beta_simp}
  dL(\tau) =
\left(
      D L^{\beta-1}
      -
      \gamma \frac{\alpha-\beta}{\alpha-\beta+2} L^{\alpha}
 \right)
       d\tau
      +
      \sqrt{D \, L^\beta} \ d W_\tau
\end{align}
 and force driven ($\alpha < \beta$)
 \begin{align}
\label{sup_tr_lv_tsa_sn_multN_alpha_sm_beta_simp}
  dL(\tau) =
  \left(
    \gamma L^\alpha
      -  D (\alpha-\beta) L^{\beta-1} 
 \right)
       d\tau
      +
      \sqrt{D \, L^\beta} \ d W_\tau
\end{align}
case. Note that the force driven case additionally requires to assume that $D$ is small whenever $-1<\alpha-\beta<0$ holds. Details of both expansions are provided in the supplementary information.

\begin{figure}[ht]
\centerline{\includegraphics[width=1.0\linewidth]
{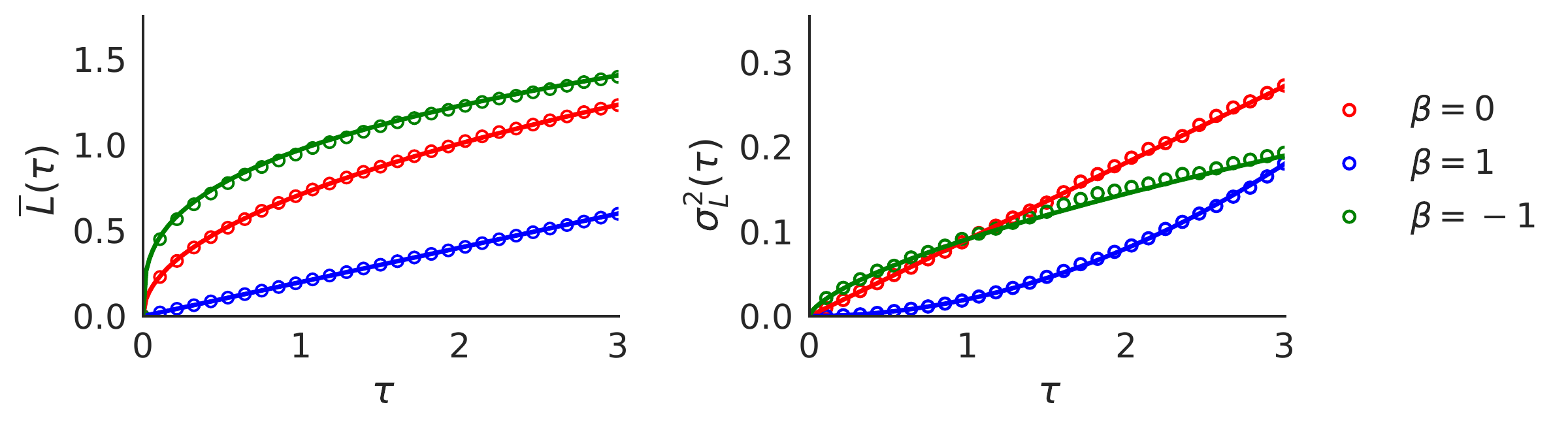}}
\caption{
\textbf{TSA ensembles with dominant multiplicative noise are distinguishable in their mean and variance.} Comparison of forward (circles) and exact (lines) multiplicative noise driven TSA dynamics with $f(L)=0$ and $D(L)=D L^\beta$ with $\beta=-1,0,1$. Shown are  the mean \textbf{(Left)} and variance \textbf{(Right)} for all three cases. The statistics of the forward dynamics are based on 20000 trajectories that start at $\widehat{L}_0 = 20$, to effectively get rid of the initial conditions. The exact analytic expression of reverse time TSA mean and variance are taken from Eq.~\eqref{MultNoiseDriven_mean} and Eq.~\eqref{MultNoiseDriven_var}. Parameters were chosen as $\gamma=1$ and $D=0.2$.
}\label{MultNoiseDriveTSAx3}
\end{figure}

In the noise driven case  ($\alpha \ge \beta$), the dynamics are dominated by a power law term that depends on the diffusion constant. For example, in the simple case of multiplicative noise with $\beta = 1$, a constant drift term of size $D$ appears which drives the reverse time dynamics away from the target state.  In general, the $D$ dependence of the leading order power law in Eq.~\eqref{sup_tr_lv_tsa_sn_multN_alpha_ge_beta_simp} suggest, that the underlying deterministic force laws are hardly distinguishable in noise driven dynamics close to the target state. We therefore exclusively use the power law term of lowest order which allows us to solve Eq.~\eqref{sup_tr_lv_tsa_sn_multN_alpha_ge_beta_simp} ($\gamma$ = 0) exactly 
(see supplementary information). This yields the mean 
\begin{align}
\label{MultNoiseDriven_mean}
\overline{L}(\tau)
=
 \frac{2^{\frac{\beta -1}{\beta -2}} (2-\beta )^{-\frac{2}{\beta -2}} \Gamma \left(-\frac{2}{\beta -2}\right) }{\Gamma \left(\frac{1}{2-\beta }\right)}
 (D\tau)^{\frac{1}{2-\beta }}
\end{align}
and the
variance
\begin{align}
\label{MultNoiseDriven_var}
\sigma_L^2(\tau)
=
&(2-\beta )^{-\frac{4}{\beta -2}} \Bigg(3\cdot 4^{\frac{1}{\beta -2}}
   \frac{\Gamma \left(-\frac{3}{\beta -2}\right)}{\Gamma \left(\frac{1}{2-\beta }\right)}
   \notag\\
   &-
   4^{\frac{\beta -1}{\beta -2}} 
   \frac{\Gamma \left(-\frac{2}{\beta-2}\right)^2}{\Gamma \left(\frac{1}{2-\beta }\right)^2}
   \Bigg)
   (D \tau)^{\frac{1}{1-\beta/2}}
   \ .
\end{align}
Disregarding the lengthy constants, both mean and variance follow a simple power law in reverse time $\tau$. For constant noise ($\beta=0$), we recover our previous result in \cite{lenner2023reversetime}. The mean increases as a square root of the time $\tau$, the variance increases linearly with time. In Fig.~\ref{MultNoiseDriveTSAx3} we compare the three cases $\beta=-1,0,1$. We find excellent agreement of our theory with the corresponding simulations, which were performed in forward time and subsequently aligned to the target state. Interestingly, all three cases ($\beta=-1,0,1$) show a clearly distinguishable behavior in their respective mean and variance. Therefore, this suggests to use the mean and variance of the TSA noise-driven dynamics for system identification for inference from experimental data.

In the force driven case  ($\alpha<\beta$) the leading order term of the reverse-time Langevin Eq.~\eqref{sup_tr_lv_tsa_sn_multN_alpha_sm_beta_simp} is the sign inverted forward force term. The diffusion constant $D$ dependent term serves as higher order correction. To obtain closed form expressions for mean, variance and covariance, we expand Eq.~\eqref{sup_tr_lv_tsa_sn_multN_alpha_sm_beta_simp} in leading orders of $D$ (see supplementary information for details). Using this small noise expansion, we determined the mean 
\begin{align}
\label{sup_mean_sn_mult}
 &\overline{L}(\tau)
 =
   ((1-\alpha ) \gamma  t)^{\frac{1}{1-\alpha }}
   \notag\\
&+
\frac{D}{\gamma}
   \frac{ \left(7 \alpha ^2-\alpha  (8 \beta +3)+2 \beta  (\beta +1)\right) ((1-\alpha) \gamma \tau)^{\frac{\alpha -\beta }{\alpha -1}}}{2 (2 \alpha -\beta ) (3 \alpha -\beta
   -1)}
\end{align}
and variance
\begin{align}
\label{sup_var_sn_mult}
 \sigma_L^2(\tau)
&=
 \frac{D}{\gamma}
 \frac{( (1 -\alpha) \gamma \tau)^{\frac{1-\alpha + \beta }{1-\alpha }}}{
 1 - 3 \alpha + \beta}
\end{align}
and two time covariance
\begin{align}
\label{sup_cov_sn_mult}
C&(\tau,\tau') =
\notag\\
=
&\begin{cases}
  D
  \frac{((1-\alpha ) \gamma  \tau')^{-\frac{\alpha }{\alpha -1}} ((1-\alpha ) \gamma 
   \tau)^{\frac{-2 \alpha +\beta +1}{1-\alpha }}}{\gamma  (1 -3 \alpha +\beta)}
   &\ \mathrm{for} \ \tau < \tau'
   \\
   D
  \frac{((1-\alpha ) \gamma  \tau)^{-\frac{\alpha }{\alpha -1}} ((1-\alpha ) \gamma 
   \tau')^{\frac{-2 \alpha +\beta +1}{1-\alpha }}}{\gamma  (1 -3 \alpha +\beta)}
   &\ \mathrm{for} \ \tau > \tau' 
   \\
   \sigma_L^2(\tau)
   &\ \mathrm{for} \ \tau = \tau'
   \ .
\end{cases}
\end{align}
up to order $D$.
\begin{figure}
\centering
\includegraphics[width=1.0\linewidth]
{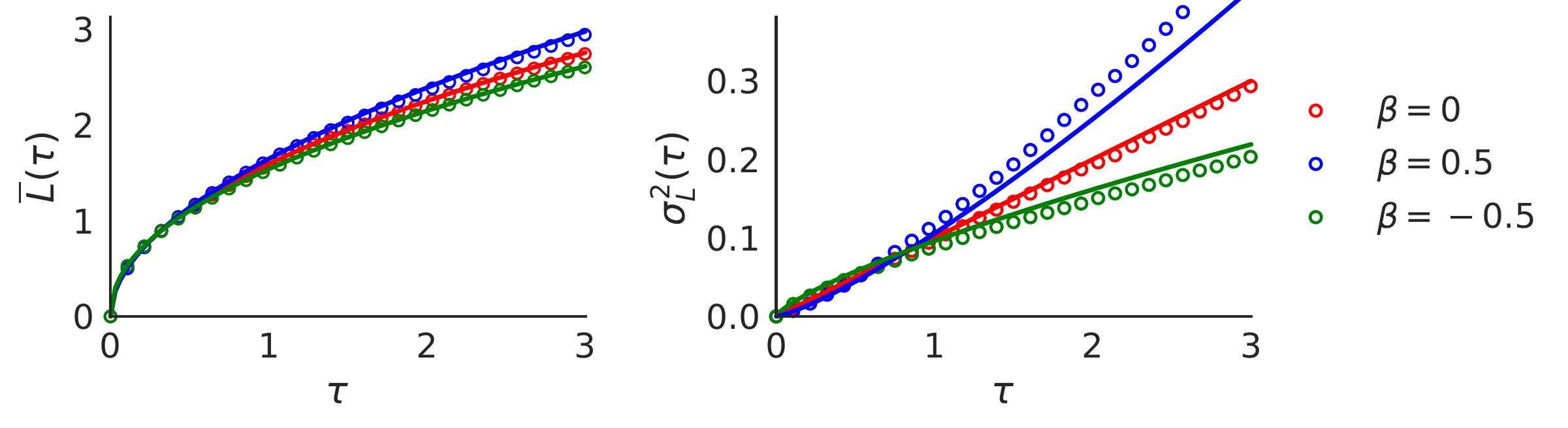}
\caption{\textbf{Force dominated TSA dynamics with multiplicative noise are distinguishable in their variance.}
Comparison of forward (circles) and small noise (lines) TSA expression for dynamics with $f(L)=-\frac{\gamma}{L}$ and multiplicative noise $D(L)=D L^\beta$ with $\beta=0,0.5,-0.5$. Shown are the mean \textbf{(Left)} and variance \textbf{(Right)} for all three cases. For the mean the influence of $\beta$ is marginal. For the variance $\beta>0$ leads to an up sloping, $\beta<0$ to a down sloping, and $\beta=0$ to a linear variance. The statistics of the forward dynamics are based on 20000 trajectories that start at $\widehat{L}_0=20$. The analytic expressions for the mean and variance are stated in Eq.~\eqref{sup_mean_sn_mult} and Eq.~\eqref{sup_var_sn_mult}. The parameters are $\gamma=1$ and $D=0.2$.
}
\label{fig:SN_mean_var}
\end{figure}

\begin{figure}
\centering
\includegraphics[width=.9\linewidth]
{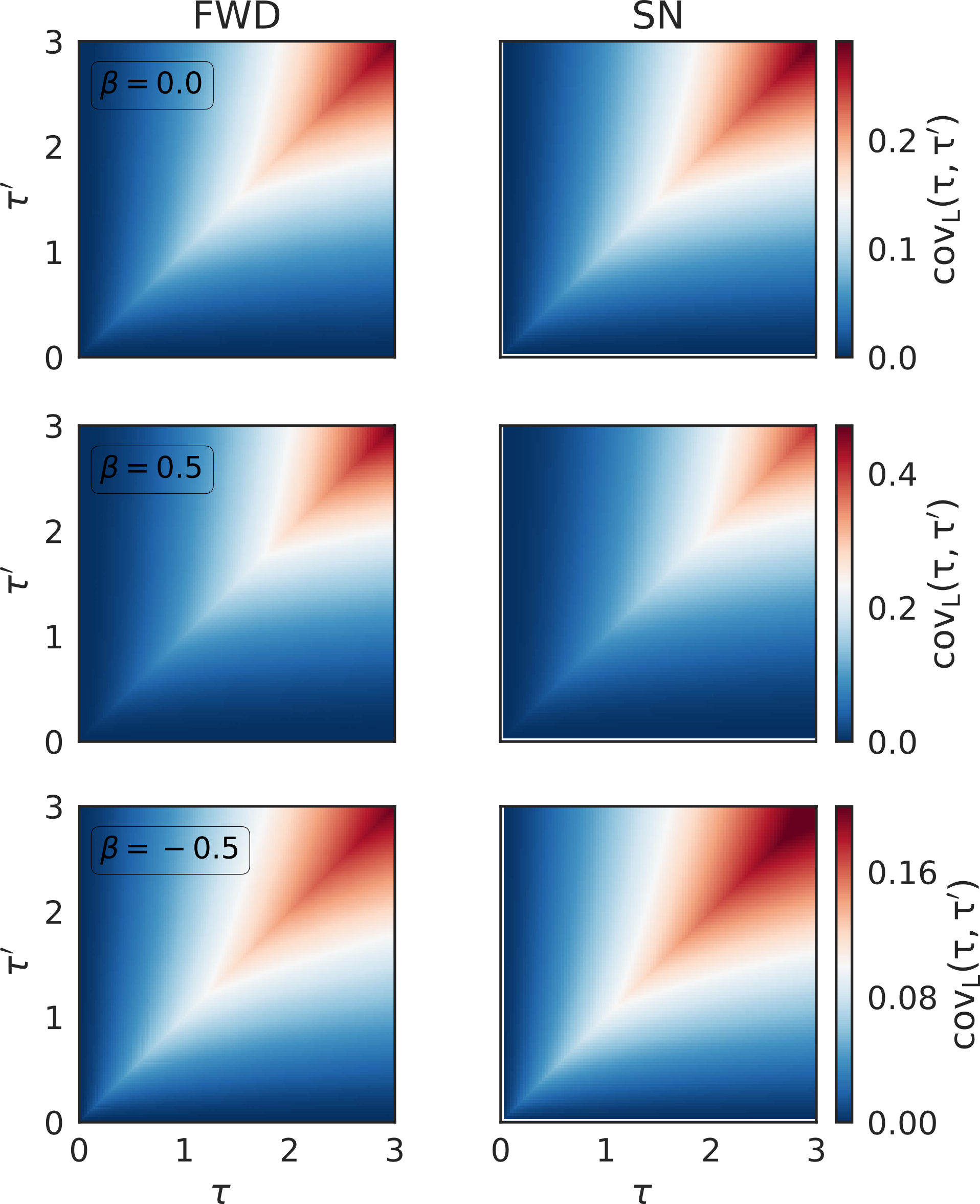}
\caption{
\textbf{The two-time covariance for force driven dynamics with multiplicative noise can be predicted from a small noise expression.} Comparison of the forward \textbf{(FWD)} and reverse time TSA small noise \textbf{(SN)} covariance for $f(L)=-\frac{\gamma}{L}$ and multiplicative noise $D(L)=D L^\beta$, with $\beta=0,0.5,-0.5$. 
The statistics of the forward dynamics are based on 20000 trajectories that start at $\widehat{L}_0=20$. The small noise covariance is stated in Eq.~\eqref{sup_cov_sn_mult}. The used parameters are $\gamma=1$ and $D=0.2$.
}
\label{fig:SN_cov}
\end{figure}

The above noted small noise results allow us to make some general observations on the behavior of force driven dynamics close to target states. Most telling about the functional characteristics of state dependent noise is the variance.
Only for $\beta =0$ the variance of the reverse time ensemble grows linearly with time. For $\beta<0$, the variance is concave close to the target state, for $\beta>0$ it is convex. We show a concrete example for all three cases in Fig.~\ref{fig:SN_mean_var}. Ensembles generated by forward simulation and subsequent target state alignment are in good agreement with the analytically obtained small noise expression of the variance Eq.~\eqref{sup_var_sn_mult}.

The mean of force driven dynamics is slightly more complicated to characterize. Dependent on the detailed parameter settings, the higher order term in the small noise expansion of the force driven mean can change from concave to convex (Eq.~\eqref{sup_mean_sn_mult}). Our small noise approximation is perfectly capable to resolve such small changes. We show this for three cases in Fig.~\ref{fig:SN_mean_var}. Changes in $\beta$ are perfectly resolvable with our higher order correction to the deterministic baseline.

The two-time covariance of force driven dynamics allows for a very visual differentiation between different multiplicative noise regimes. We show three different noise regimes in Fig.~\ref{fig:SN_cov}. For each case, our small noise approximation of the covariance (Eq.~\eqref{sup_cov_sn_mult}) is in good agreement with the covariance obtained from forward simulated and subsequently target state aligned trajectories.

\section{DISCUSSION}
Typical physics problems prime us to expect dynamical systems problems as initial value problems. Often we can indeed either theoretically or experimentally prepare the system in a well defined initial state and then propagate the dynamics in forward time. In stochastic dynamics we additionally need to account for random forces that perturb the dynamics. But also in stochastic dynamics we already know best how to analyze initial value problems.
The temporal evolution of stochastic processes is canonically described by Langevin equations if the focus lies on the generation of individual sample paths and the forward Fokker Planck equation to describe the evolution of the ensemble \cite{gardiner1985handbook}. Both can be used to describe initial value problems, that is dynamics starting from an initial distribution. In living systems, however, such initial value problems are rare. The biological dynamics we studied above are much better characterized as "final value"
problems reflecting that functional demands require canalization towards target states \cite{sha2003hysteresis,pomerening2003building,coudreuse2010driving,rata2018two,schwarz2018precise,domingo2011switches,nachman2007dissecting, pardee1974restriction,ahrends2014controlling,maamar2007noise,xiong2003positive,losick2008stochasticity,balazsi2011cellular,hanes1996neural,hanks2015distinct,ratcliff2016diffusion,ratcliff2008diffusion,brunton2013rats,hanks2015distinct,churchland2011variance,roitman2002response}. 

Interventions to artificially introduce initial conditions are typically practically inconceivable and would in fact not interrogate the naturally occurring dynamics. We therefore suggested in \cite{lenner2023reversetime} to analyze end state directed dynamics in their natural frame of reference, that is, in reverse time as a target state aligned ensemble. In our previous study, we showed how such target state aligned (TSA) ensembles can be analyzed for the simpler case of state-independent noise. In the current paper we extend this framework to include general position dependent diffusion, that is multiplicative noise. In this setting, we find target state directed dynamics with much richer behavior close to target states. Including multiplicative noise also expands the applicability of our approach to a new range of problems in finance, neuroscience, chemistry, evolution and cell biology \cite{LIMA2018, laing2009stochastic, gillespie2000chemical, wright1931evolution, fisher1915evolution}. 

In the case of constant noise, all dynamics with a force vector pointing towards the target state will eventually reach the target state. For dynamics with multiplicative noise this is not any longer guaranteed. We determined the class of dynamics which on average reach the target state in finite time. We determine the universal form of power laws close to the target states and derive expressions for mean, variance and covariance –both in the noise dominated or force dominated limit. These expressions  provide an intuition for possible underlying dynamics in TSA ensembles and form a solid foundation for the data-driven identification of target state directed dynamical systems. Target state alignment is a technique which allows to analyze directional stochastic dynamics with uncontrollable initial conditions. Intriguingly, target states alignment however creates pseudo-forces that have to be separated from the true underlying forward dynamics. Our theory of target state aligned ensembles, allows us to map out the range of inferable TSA dynamics with quantitative precision and principled mathematical controls.




\begin{acknowledgments}
We thank Erik Schultheis and the Wolf group for stimulating discussions and proofreading. This work was supported by the German Research Foundation (Deutsche Forschungsgemeinschaft, DFG) through FOR 1756, SPP 1782, SFB 1528, SFB 889, SFB 1286, SPP 2205, DFG 436260547 in relation to NeuroNex (National Science Foundation 2015276) \& under Germany’s Excellence Strategy - EXC 2067/1- 390729940; by the Leibniz Association (project K265/2019); and by the Niedersächsisches Vorab of the VolkswagenStiftung through the Göttingen Campus Institute for Dynamics of Biological Networks.
\end{acknowledgments}


\bibliography{literature}
\end{document}



\title{Supplementary Information - Reverse-time analysis and boundary classification of directional biological dynamics with multiplicative noise}

\author{Nicolas Lenner}
 \altaffiliation[Currently at ]{Simons Center for Systems Biology, School of Natural Sciences, Institute for Advanced Study, Princeton, New Jersey, USA.}
  \email{Lenner@ias.edu}
 \affiliation{Max Planck Institute for Dynamics and Self-Organization, Göttingen, Germany}

\author{Matthias Häring}
 \affiliation{Max Planck Institute for Dynamics and Self-Organization, Göttingen, Germany}
\affiliation{Göttingen Campus Institute for Dynamics of Biological Networks, University of Göttingen, Göttingen, Germany}

\author{Stephan Eule}%
\affiliation{Max Planck Institute for Dynamics and Self-Organization, Göttingen, Germany}
\affiliation{German Primate Center—Leibniz Institute for Primate Research, Goettingen, Germany}

\author{Jörg Großhans}
\affiliation{Department of Biology, Philipps University Marburg, Marburg, Germany}
\affiliation{Göttingen Campus Institute for Dynamics of Biological Networks, University of Göttingen, Göttingen, Germany}

\author{Fred Wolf}
 \email{Fred.Wolf@ds.mpg.de}
\affiliation{Max Planck Institute for Dynamics and Self-Organization, Göttingen, Germany}
\affiliation{Göttingen Campus Institute for Dynamics of Biological Networks, University of Göttingen, Göttingen, Germany}
\affiliation{Max Planck Institute for Multidisciplinary Sciences, Göttingen, Germany}
\affiliation{Institute for the Dynamics of Complex Systems, University of Göttingen, Göttingen, Germany}
\affiliation{Center for Biostructural Imaging of Neurodegeneration, Göttingen, Germany}
\affiliation{Bernstein Center for Computational Neuroscience Göttingen, Göttingen, Germany}


\maketitle


%
%
%
%
%
%
%
%

\newpage
\tableofcontents
\newpage


\section{The TSA ensemble for dynamics with multiplicative noise}
In forward time, the dynamics of sample paths are governed by either the Langevin equation or equivalently the Fokker-Planck equation (FPE). For the following derivations we use the forward FPE
\begin{align}
 \label{sup_fwd_fp_mult}
 \frac{\partial}{\partial t} 
P(\widehat{L},t|\widehat{L}_0,t_0)
 =
    - \frac{\partial}{\partial \widehat{L}} f(\widehat{L}) \,
P(\widehat{L},t|\widehat{L}_0,t_0)
    + \frac{1}{2} \frac{\partial^2}{\partial \widehat{L}^2} 
 D(\widehat{L})   
P(\widehat{L},t|\widehat{L}_0,t_0)
\end{align}
and its adjoint, the backward FPE
\begin{align}
 \label{sup_bwd_fp_mult}
  \frac{\partial}{\partial t} P(\widehat{L}_f,t_f | \widehat{L},t) 
  =
    - f(\widehat{L}) \frac{\partial}{\partial \widehat{L}} P(\widehat{L}_f,t_f | 
\widehat{L},t)
    - \frac{1}{2}  
    D(\widehat{L}) 
      \frac{\partial^2 }{\partial \widehat{L}^2} P(\widehat{L}_f,t_f | 
\widehat{L},t) 
\ .
\end{align}
We here chose the Ito interpretation to define these two Fokker-Planck equations \cite{van1981ito}. For the forward Fokker-Planck equation this implies that the diffusion ``constant'' $D(\widehat{L})$ ends up inside of the second derivative, for the backward Fokker-Planck equation the diffusion "constant" is evaluated outside of the derivative. 
For a thorough discussion which interpretation should be chosen under different experimental conditions and how to transform between different interpretations we refer to van Kampen \cite{van1981ito} and Sokolov \cite{sokolov2010ito}. Throughout this text we use the Ito-interpretation where the stochastic forcing is always evaluated at the onset of each time increment \cite{gardiner1985handbook}. The subsequent sections are organized as follows. We start with the derivation of the theory, briefly touch upon the construction of TSA dynamics close to the target state and conclude with a small noise expansion of TSA dynamics with power law like force and noise terms.

\subsection{Construction of the TSA ensemble}
The construction idea of TSA ensembles is based on (i) dividing a complete ensemble of sample paths, which all have reached a target state (TS) into sub-ensembles of lifetime $T_i$, (ii) reverting these sub-ensembles and (iii) aligning them to the new initial i.e.~the former TS. In general, a sub-ensemble is mathematically defined as a bridge process \cite{majumdar2015effective}
\begin{align}
\label{sup_bridge_trans_nobound}
P(\widehat{L},t|\widehat{L}_f,t_f,\widehat{L}_0,t_0) 
= 
\frac{
P(\widehat{L}_f,t_t|\widehat{L},t) P(\widehat{L},t|\widehat{L}_0,t_0)
}{
P(\widehat{L}_f,t_f|\widehat{L}_0,t_0) 
}
\ ,
\end{align}
which characterizes transitions from an initial state to an intermediate state and then to the final state. The individual transition probability densities all obey the FPE. Bridge processes are thus dynamics generated by a FPE with selected initial and final condition and subsequently normalized to this selection. For dynamics which, as in our case, terminate at an absorbing boundary, i.e.~ target state, a problem arises. Absorbing boundary conditions are typically implemented by setting $P(\widehat{L},t|\widehat{L}_0,t_0)$ and $P(\widehat{L}_f,t_t|\widehat{L},t)$ to zero at the boundary. As we are interested in the probability flux into the bounary at time $t_f$, $P(\widehat{L}_f,t_t|\widehat{L},t)$ must be replaced by the hitting time distribution, defined as
\begin{align}
\label{eq:def_hit_dist}
\rho_{\widehat{L}_\mathrm{ts}}(t_f|\widehat{L},t) 
=
- \frac{\partial}{\partial t_f}
\int_{\widehat{L}_\mathrm{ts}}^\infty dL_f \ P(\widehat{L}_f,t_f|\widehat{L},t)
\ .
\end{align}
Bridge processes ending at the target state then read
\begin{align}
\label{sup_bridge_trans_bound}
R_{\widehat{L}_\mathrm{ts}}(\widehat{L},t|\widehat{L}_0,t_0;t_f)
:=
 \frac{\rho_{\widehat{L}_\mathrm{ts}}(t_f|\widehat{L},t)
        P(\widehat{L},t|\widehat{L}_0,t_0)
 }{\rho_{\widehat{L}_\mathrm{ts}}(t_f|\widehat{L}_0,t_0)}
 \ ,
\end{align}
where we adapted the normalization according to Lenner et.al.~\cite{lenner2023reversetime}.

For the construction of the TSA ensemble, we need to average over all contributing sub-ensembles, i.e.~bridge processes. Their respective weight to the complete ensemble is given by the hitting-time distribution. In forward time the TSA ensemble then reads
\begin{align}
\label{sup_prob_full_tr_ens}
R_{\widehat{L}_\mathrm{ts}}(\widehat{L},t;t_f|\widehat{L}_0)
=
\int_{-\infty}^t
dt_0
\rho_{\widehat{L}_\mathrm{ts}}(t_f|\widehat{L}_0,t_0)
R_{\widehat{L}_\mathrm{ts}}(\widehat{L},t|\widehat{L}_0,t_0;t_f)
\ ,
\end{align}
where all sub-ensembles are aligned to reach the target state at the same reference time $t_f$.
The upper integral limit ensures that the averaging only includes sample paths which already exist at time $t$. This expression can be generalized to arbitrary initial distributions by averaging over $R_{\widehat{L}_\mathrm{ts}}(\widehat{L},t;t_f|\widehat{L}_0)$ with respect to  $P^\mathrm{in}(\widehat{L}_0)$ to obtain
\begin{align}
\label{sup_prob_full_tr_ens_initDist}
R_{\widehat{L}_\mathrm{ts}}(\widehat{L},t;t_f)
&=
\int_{\widehat{L}_\mathrm{ts}}^\infty
d\widehat{L}_0
P^\mathrm{in}(\widehat{L}_0)
\int_{-\infty}^t
dt_0
\rho_{\widehat{L}_\mathrm{ts}}(t_f|\widehat{L}_0,t_0)
R_{\widehat{L}_\mathrm{ts}}(\widehat{L},t|\widehat{L}_0,t_0;t_f)
\ .
\intertext{Using the definition of bridge processes ending at a target state Eq.~\eqref{sup_bridge_trans_bound} this expression simplifies to}
\label{sup_prob_full_tr_ens_initDist_2nd}
&=
\int_{\widehat{L}_\mathrm{ts}}^\infty
d\widehat{L}_0
P^\mathrm{in}(\widehat{L}_0)
\int_{-\infty}^t
dt_0
\rho_{\widehat{L}_\mathrm{ts}}(t_f|\widehat{L},t) \ P(\widehat{L},t|\widehat{L}_0,t_0)
\ .
\end{align}
For the following derivation we use the definition of the TSA ensemble in forward time. Only in the final step we well shift to the time to completion framework $\tau=t_f -t$ to yield the reverse time TSA ensemble
\begin{align}
\label{sup_rtsa_dist}
 R_{L_\mathrm{ts}}(L,\tau;\tau_0)
=
\int_{L_\mathrm{ts}}^\infty
dL_f
P^\mathrm{in}(L_f)
\int_\tau^\infty d\tau_f
\rho_{L_\mathrm{ts}}(\tau_0|L,\tau)
\
P(L,\tau|L_f,\tau_f)
 \ .
\end{align}

The temporal evolution of $R_{L_\mathrm{ts}}(L,\tau;\tau_0)$ is obtained from a time $t$ derivative of the forward in time expression $R_{\widehat{L}_\mathrm{ts}}(\widehat{L},t;t_f)$. Using the Leibnitz rule and evaluating terms we find
\begin{align}
\label{sup_tempEvo_TSA_ens_applLeibnitz}
\frac{\partial}{\partial t}
R_{\widehat{L}_\mathrm{ts}}(\widehat{L},t;t_f)
=
\rho_{\widehat{L}_\mathrm{ts}}(t_f|\widehat{L},t) 
P^\mathrm{in}(\widehat{L})
+
\int_{\widehat{L}_\mathrm{ts}}^\infty
d\widehat{L}_0
P^\mathrm{in}(\widehat{L}_0)
\left(
\int_{-\infty}^t dt_0 
 \frac{\partial}{\partial t}
 \left(
 \rho_{\widehat{L}_\mathrm{ts}}(t_f|\widehat{L},t) \ P(\widehat{L},t|\widehat{L}_0,t_0)
\ 
 \right)
\right)
\ .
\end{align}
This is our first intermediate result. In the following steps, we derive a Fokker-Planck expression for the evolution of the product of the hitting time distribution and the transition probability inside the integral, which, according to our notation is defined as
\begin{align}
\label{sup_prod_of_hit_and_transProb}
 R_{\widehat{L}_\mathrm{ts}}(\widehat{L},t;t_f|\widehat{L}_0,t_0)
 =
 \rho_{\widehat{L}_\mathrm{ts}}(t_f|\widehat{L},t) \ P(\widehat{L},t|\widehat{L}_0,t_0)
\ .
\end{align}
We apply the chain rule and use that $P(\widehat{L},t|\widehat{L}_0,t_0)$ obeys the forward FPE and 
$\rho_{\widehat{L}_\mathrm{ts}}(t_f|\widehat{L},t)$ the backward FPE to write
\begin{align} 
\label{joint_fpPlugIn_mult}
 - \frac{\partial 
 R_{\widehat{L}_\mathrm{ts}}(\widehat{L},t;t_f|\widehat{L}_0,t_0)
}{\partial t}  
 &= P(\widehat{L},t|\widehat{L}_0,t_0)
    \left[ 
    f(\widehat{L}) \frac{\partial}{\partial \widehat{L}} 
    \rho_{\widehat{L}_\mathrm{ts}}(t_f|\widehat{L},t)
    + \frac{1}{2} D(\widehat{L})  
      \frac{\partial^2}{\partial \widehat{L}^2} 
      \rho_{\widehat{L}_\mathrm{ts}}(t_f|\widehat{L},t)
    \right]
 \notag \\
 &\ \ \ + 
 \rho_{\widehat{L}_\mathrm{ts}}(t_f|\widehat{L},t)
    \left[
    \frac{\partial}{\partial \widehat{L}} f(\widehat{L}) \, P(\widehat{L},t|\widehat{L}_0,t_0)
    - \frac{1}{2} \frac{\partial^2}{\partial \widehat{L}^2}  D(\widehat{L})   
    P(\widehat{L},t|\widehat{L}_0,t_0)
    \right] \quad .
\end{align}
In the next step we seek to combine the transition probability $P(\widehat{L},t|\widehat{L}_0,t_0)$, the hitting time distribution $\rho_{\widehat{L}_\mathrm{ts}}(t_f|\widehat{L},t)$ and the diffusion ``constant'' $D(\widehat{L})$ into the form of the diffusion term of a Fokker-Planck equation in Ito-representation.
Using the chain-rule once on the drift terms and twice on the diffusive terms we find
\begin{align}
\label{sup_rt_fp_full_dist_subst_mult}
 - \frac{\partial 
 R_{\widehat{L}_\mathrm{ts}}(\widehat{L},t;t_f|\widehat{L}_0,t_0)
 }{\partial t}  
 &=
     \frac{\partial}{\partial \widehat{L}}
    \left[
      f(\widehat{L}) R_{\widehat{L}_\mathrm{ts}}(\widehat{L},t;t_f|\widehat{L}_0,t_0)
      - \ \rho_{\widehat{L}_\mathrm{ts}}(t_f|\widehat{L},t)
      \frac{\partial}{\partial 
\widehat{L}} D(\widehat{L}) P(\widehat{L},t|\widehat{L}_0,t_0)
    \right] 
\notag \\
    &+ \frac{1}{2} \frac{\partial^2}{\partial \widehat{L}^2} 
    D(\widehat{L})
    R_{\widehat{L}_\mathrm{ts}}(\widehat{L},t;t_f|\widehat{L}_0,t_0)
\ .
\end{align}
Note that the diffusion term occurs with a minus sign compared to the time derivative. This will change when rewriting the time in time to completion $\tau$.
Substituting this expression back into the evolution equation of the TSA ensemble Eq.~\eqref{sup_tempEvo_TSA_ens_applLeibnitz} and 
using the definition of $R_{\widehat{L}_\mathrm{ts}}(\widehat{L},t;t_f)$
in Eq.~\eqref{sup_prob_full_tr_ens_initDist_2nd} together with Eq.~\eqref{sup_prod_of_hit_and_transProb}, we arrive at
\begin{align}
\label{sup_1st_FP_for_full_TSA}
\frac{\partial}{\partial t}
R_{\widehat{L}_\mathrm{ts}}(\widehat{L},t;t_f)
=
\
 &\rho_{\widehat{L}_\mathrm{ts}}(t_f|\widehat{L},t) P^\mathrm{in}(\widehat{L})
\notag \\
     &-
     \frac{\partial}{\partial \widehat{L}}
    \left[
      f(\widehat{L}) 
      R_{\widehat{L}_\mathrm{ts}}(\widehat{L},t;t_f)
      -  \,
\rho_{\widehat{L}_\mathrm{ts}}(t_f|\widehat{L},t) \frac{\partial}{\partial 
\widehat{L}}
D(\widehat{L})
\int_{\widehat{L}_\mathrm{ts}}^\infty 
d\widehat{L}_0
P^\mathrm{in}(\widehat{L}_0)
      \int_{-\infty}^t dt_0
P(\widehat{L},t|\widehat{L}_0,t_0)
    \right] 
\notag \\    
    &- 
    \frac{1}{2} \frac{\partial^2}{\partial \widehat{L}^2} 
    D(\widehat{L})
 R_{\widehat{L}_\mathrm{ts}}(\widehat{L},t;t_f)
\ .
\intertext{
We next multiply the remaining $\rho_{\widehat{L}_\mathrm{ts}}(t_f|\widehat{L},t)$ in the advection part of Eq.~\eqref{sup_1st_FP_for_full_TSA} by 
$
D(\widehat{L})
\int_{\widehat{L}_\mathrm{ts}}^\infty 
d\widehat{L}_0
P^\mathrm{in}(\widehat{L}_0)
      \int_{-\infty}^t dt_0
P(\widehat{L},t|\widehat{L}_0,t_0)
$
both in nominator and denominator to obtain a function of $R_{\widehat{L}_\mathrm{ts}}(\widehat{L},t;t_f)$ again using Eq.~\eqref{sup_prob_full_tr_ens_initDist_2nd}. 
Joining the remaining terms of this expression into the derivative of a logarithm. We find
}
\label{sup_full_tsa_pfwd}
&=
\rho_{\widehat{L}_\mathrm{ts}}(t_f|\widehat{L},t) P^\mathrm{in}(\widehat{L})
\notag \\
     &-
     \frac{\partial}{\partial \widehat{L}}
    \left[
      f(\widehat{L}) 
      - 
D(\widehat{L})
\frac{\partial}{\partial \widehat{L}}
\log 
\left(
D(\widehat{L})
\int_{\widehat{L}_\mathrm{ts}}^\infty 
d\widehat{L}_0
P^\mathrm{in}(\widehat{L}_0)
      \int_{-\infty}^t dt_0
P(\widehat{L},t|\widehat{L}_0,t_0)
\right)
    \right] 
    R_{\widehat{L}_\mathrm{ts}}(\widehat{L},t;t_f)
\notag \\    
    &- 
    \frac{1}{2} \frac{\partial^2}{\partial \widehat{L}^2} 
D(\widehat{L})
 R_{\widehat{L}_\mathrm{ts}}(\widehat{L},t;t_f)
 \ .
\end{align}
%
%
%
%
%
%
We next seek a more accessible expression for the double integral inside the logarithm.
For most cases no analytic solution for $P(\widehat{L},t|\widehat{L}_0,t_0)$ is available and the associated forward FPE can not be solved analytically. Instead of searching for an solution for $P(\widehat{L},t|\widehat{L}_0,t_0)$ we solve the forward FPE after integrating out the dependence on the initial time $t_0$. In particular we find
\begin{align}
 \int_{-\infty}^t dt_0
 \frac{\partial}{\partial t}  P(\widehat{L},t|\widehat{L}_0,t_0)
 = 
 - \int_{\infty}^0 ds
 \frac{\partial}{\partial s}  P(\widehat{L},s|\widehat{L}_0,0) = -P(\widehat{L},0|\widehat{L}_0,0) 
 =
 -\delta(\widehat{L}-\widehat{L}_0) 
\end{align}
We then define the non-equilibrium stationary density
\begin{align}
\label{sup_cfb_density}
Q(\widehat{L})
:=
\lambda
\int_{\widehat{L}_\mathrm{ts}}^{\infty } d\widehat{L}_0 \; P^\mathrm{in}(\widehat{L}_0)
 \int_{-\infty}^t dt_0
P(\widehat{L},t|\widehat{L}_0,t_0)
\ ,
\end{align}
where $\lambda$ ensures the normalization.
Under these two substitutions the forward FPE simplifies to the ordinary differential equation
\begin{align}
\label{ness_ode_mult}
 - 
 \lambda
 P^\mathrm{in}(\widehat{L}) 
 &= 
 - \frac{\partial}{\partial \widehat{L}} \left( f(\widehat{L}) Q(\widehat{L}) \right)
 + \frac{1}{2} \frac{\partial^2}{\partial \widehat{L}^2} D(\widehat{L}) Q(\widehat{L})
 \notag \\
 Q(\widehat{L}_\mathrm{ts})&=0
 .
\end{align}
The normalization factor 
$\lambda = \frac{1}{2} \frac{\partial}{\partial L} D(\widehat{L}_\mathrm{ts}) Q(\widehat{L}_\mathrm{ts})$ is obtained by integration over the full interval 
$[\widehat{L}_\mathrm{ts},\infty]$ with 
$\int_{\widehat{L}_\mathrm{ts}}^{\infty} P^{\mathrm{in}}(\widehat{L'}) d\widehat{L'} =1$.
Solved in a different context, the solution of Eq.~\eqref{ness_ode_mult} is known \cite{zhang2011reconstructing}.
Integrating over the interval $\widehat{L}_\mathrm{ts}$ to  $\widehat{L}$ one arrives at
\begin{equation}
 -
 f(\widehat{L}) Q(\widehat{L})
 +
 \frac{1}{2}
 \frac{\partial}{\partial \widehat{L}}
D(\widehat{L})
  Q(\widehat{L})
 =
 \lambda
 \left(
 1
 - 
 \int_{\widehat{L}_\mathrm{ts}}^{\widehat{L}} 
 P^{\mathrm{in}}(\widehat{L'}) d\widehat{L'}
 \right)
\ .
 \end{equation}
Combining the terms on the left hand side into one expression then yields
\begin{align}
 \frac{1}{2} 
  e^{
  \int^{\widehat{L}}
 \frac{2 f(\widehat{L'})}{D(\widehat{L'})} d\widehat{L'}
 }
 \frac{\partial}{\partial \widehat{L}}
 \left(
 D(\widehat{L})
 Q(\widehat{L})
 e^{
  -
  \int^{\widehat{L}}
 \frac{2 f(\widehat{L'})}{D(\widehat{L'})} d\widehat{L'}
 }
 \right)
 &=
 \lambda
 \left(
 1
 -
 \int_{\widehat{L}_\mathrm{ts}}^{\widehat{L}} P^{\mathrm{in}}(\widehat{L'}) d\widehat{L'}
 \right) 
 \ .
\end{align}
Again integrating from $\widehat{L}_\mathrm{ts}$ to  $\widehat{L}$ and solving for $Q(\widehat{L})$ yields the final seeked expression
\begin{align}
\label{sup_cfb_density_final}
 Q(\widehat{L})
 &=
 \frac{2\lambda}{D(\widehat{L})}
 \;
   e^{
    \int^{\widehat{L}}
  \frac{2 f(\widehat{L'})}{D(\widehat{L'})} d\widehat{L'}
  }
 \int_{\widehat{L}_\mathrm{ts}}^{\widehat{L}} d\widehat{L'} \;
   e^{-
  \int^{\widehat{L'}}
 \frac{2 f(\widehat{L''})}{D(\widehat{L''})} d\widehat{L''}
 }
 \left(
 1
 -
 \int_{\widehat{L}_\mathrm{ts}}^{\widehat{L'}} P^{\mathrm{in}}(\widehat{L''}) d\widehat{L''}
 \right) 
  \ .
\end{align}
This result can now be used to rewrite the lengthy logarithmic advection term in Eq.~\eqref{sup_full_tsa_pfwd}. First we substitute the double integral inside the logarithm with $Q(\widehat{L})$ as defined in Eq.~\eqref{sup_cfb_density}. Next, we substitute $Q(\widehat{L})$ with our result from Eq.~\eqref{sup_cfb_density_final} to obtain 
\begin{align}
 D(\widehat{L}) \frac{\partial}{\partial \widehat{L}} \log D(\widehat{L}) Q(\widehat{L})
 =
 2 f(\widehat{L}) 
 +
 D(\widehat{L}) \log\left(
  \int_{\widehat{L}_\mathrm{ts}}^{\widehat{L}} d\widehat{L'} \;
   e^{-
  \int^{\widehat{L'}}
 \frac{2 f(\widehat{L''})}{D(\widehat{L''})} d\widehat{L''}
 }
 \left(
 1
 -
 \int_{\widehat{L}_\mathrm{ts}}^{\widehat{L'}} P^{\mathrm{in}}(\widehat{L''}) d\widehat{L''}
 \right) 
 \right)
 \ ,
\end{align}
for the logarithmic advection term.
We note that the $D(\widehat{L})$ terms inside the logarithm have canceled.

Substituting this expression into Eq.~\eqref{sup_full_tsa_pfwd} for the logarithmic term
and introducing the time to completion $\tau=t_f-t$,
directly leads to the reverse time TSA Fokker-Planck equation with multiplicative noise
\begin{align}
\label{eq:revt_TSA_FP_mult}
&\frac{\partial}{\partial \tau}
R_{L_\mathrm{ts}}(L,\tau;\tau_0)
  =
-
  P^\mathrm{in}(L)
 \rho_{L_\mathrm{ts}}(\tau|L) 
\notag \\
 &-
     \frac{\partial}{\partial L}
     \left(\!
    \left[
       f(L) 
      +D(L) 
      \  
      \frac{\partial}{\partial 
L}
 \log\! \left(
  \int_{L_\mathrm{ts}}^{L} dL' \;
   e^{-
  \int^{L'}
 \frac{2 f(L'')}{D(L'')} dL''
 }
 \left(
 1
 -
 \int_{L_\mathrm{ts}}^{L'} P^{\mathrm{in}}(L'') dL''
 \!
 \right)
 \!
 \right)
\!
    \right] 
     R_{L_\mathrm{ts}}(L,\tau;\tau_0)
     \!
     \right)
\notag \\    
    &+ \frac{1}{2} \frac{\partial^2}{\partial L^2}  
D(L)
R_{L_\mathrm{ts}}(L,\tau;\tau_0)
\ .
\end{align} 
Note, that the resulting ensemble is not normalized but decays with a rate proportional to the hitting time $\rho_{L_\mathrm{ts}}(\tau|L)$. This is exactly in line with the construction idea of TSA ensembles.

In the next step we associate the reverse time TSA Fokker-Planck equation with a Langevin equation. With the diffusion ``constant'' $D(L)$ inside of the second derivative and the forces inside of the first derivative, Eq.~\eqref{eq:revt_TSA_FP_mult} is clearly of Ito-Form. The corresponding Langevin equation thus reads
\begin{align}
\label{sup_tr_lv_tsa_mult}
  dL(\tau) =
\left(
      f(L) 
      +
      f^\mathcal{F}(L)
 \right)
       d\tau
      +
      \sqrt{D(L)} \ d W_\tau
     \ ,
\end{align}
where we call the second force term
\begin{align}
\label{sup_freeE_force_multNoise}
f^\mathcal{F}(L)
=
D(L) 
      \  
      \frac{\partial}{\partial 
L}
 \log\left(
  \int_{L_\mathrm{ts}}^{L} dL' \;
   e^{-
  \int^{L'}
 \frac{2 f(L'')}{D(L'')} dL''
 }
 H(L')
 \right)  
 \ ,
\end{align}
a free energy force. It accounts for the time-reversal and alignment of the TSA dynamics. The term
\begin{align}
\label{sup_H_F_dep_on_in_cond}
H(L')
=
 1
 -
 \int_{L_\mathrm{ts}}^{L'} P^{\mathrm{in}}(L'') dL''
\end{align}
denotes the dependence of the free energy force on the initial conditions of the forward dynamics. It is given as one minus the cumulative distribution of the initial distribution of the forward process.
For this Langevin equation, we represent the sink in the FPE equation with a killing measure \cite{holcman2005survival,schuss2015brownian,erban2007reactive}
\begin{align}
\label{sup_killing_measure_def}
  k(L,\tau) d\tau
 =\frac{
   P^\mathrm{in}(L)
 \rho_{L_\mathrm{ts}}(\tau_0|L,\tau)
 }{
 R_{L_\mathrm{ts}}(L,\tau;\tau_0)
 }
 d\tau
 \ .
\end{align}
This killing measure defines the fraction of sample paths that must be terminated in each timestep $\Delta \tau$. In general it is surprisingly difficult to correctly implement a killing measure for Langevin equations \cite{erban2007reactive}. In our specific case
we face the additional complication, that the killing measure is only defined on the ensemble level, that is, via the distribution of not yet terminated sample paths  $R_{L_\mathrm{ts}}(L,\tau;\tau_0)$.
This dependency on the ensemble however also allows us to implement the killing measure as a counter of trajectories that must be terminated per time step and spatial interval. To be precise, in each time step and for each spatial interval the fraction 
\begin{align}
k(L,\tau) \Delta\tau
=
\min
\left(1,
\frac{
P^\mathrm{in}(L)
\rho_{L_\mathrm{ts}}(\tau_0|L,\tau)
\Delta\tau \ \Delta L
}{
n_{\tau,\Delta L}
}
\right)
\end{align} 
of all still per bin present trajectories $n_{\tau,\Delta L}$ is terminated. We propose to implement this empirical killing measure as a Metropolis like criterion \cite{metropolis1953equation}: For values of the empirical killing measure larger one, all sample paths in the bin are terminated. For values below one, an iid random number compared to the stated fraction decides for each sample path whether it is terminated or not. If the iid random number happens to be smaller than the fraction, the sample path is terminated.

\subsection{Exactly solvable TSA dynamics}
\label{sec_Exactly solvable reverse time TSA dynamics with multiplicative noise}
The exact solution of the reverse time TSA Fokker-Planck equation with multiplicative noise is obtainable only for a few cases. We here study the case
\begin{equation}
\label{sup_fwd_sde_bessel_mult_ForceAndNoise}
 d\widehat{L}(t) = - \gamma \, dt + \sqrt{D \widehat{L}} \, dW_t
 \ .
\end{equation}
where the SDE is interpreted in Ito sense. 
For simplicity we choose $\delta$-initial conditions at $\widehat{L}_0=L_f$. The TSA reverse time SDE to be solved then reads
\begin{align}
\label{sup_tr_lv_bessel_tsa_simp_mult}
  dL(\tau) =
  \begin{cases}
\left(\gamma + D\right) \;
       d\tau
      &+ \;
      \sqrt{D L} \ d W_\tau
      \qquad \mathrm{for} \ \ L < L_f
      \\
- \gamma \;
       d\tau      
      &+ \;
      \sqrt{D L} \ d W_\tau
      \qquad \mathrm{for} \ \ L > L_f
    \ ,
  \end{cases}
\end{align}
and is
directly obtained from the definition of the reverse time TSA SDE with multiplicative noise in Eq.~\eqref{sup_tr_lv_tsa_mult}.

To construct a solution for Eq.~\eqref{sup_tr_lv_bessel_tsa_simp_mult}
we exploit that Eq.~\eqref{sup_fwd_sde_bessel_mult_ForceAndNoise} is equivalent to a Bessel process \cite{bray2000random}, as can be seen using Ito's lemma \cite{gardiner1985handbook}. Under this lemma one can show that a SDE of the form of the Bessel-process \cite{bray2000random}
\begin{align}
 dx(t) = - \frac{\gamma}{x} \, dt + \sqrt{D} \, dW_t
\end{align}
transforms to
\begin{align}
 dy(t) = \left(-2 \gamma + D \right) \, dt + \sqrt{4 D y} \, dW_t
\end{align}
using a variable transformation of the form $y=x^2$. Substituting $D=\frac{\widetilde{D}}{4}$ and $\gamma=\frac{\widetilde{\gamma}}{2} + \frac{\widetilde{D}}{8}$,
we arrive at Eq.~\eqref{sup_fwd_sde_bessel_mult_ForceAndNoise} with the ``$\sim$'' removed from display for notational unity. 
Knowing the correct variable transformation and parameter substitutions we can directly transform the solution of the TSA Bessel process, as derived in Lenner et.~\cite{lenner2023reversetime}, into the solution of Eq.~\eqref{sup_tr_lv_bessel_tsa_simp_mult} and find
\begin{align}
\label{sol_bessel_mult}
R_0(L,\tau;L_f)
=
 \begin{cases}
 \frac{4^{\frac{\gamma +D}{D}} (D \tau)^{-\frac{2 (\gamma +D)}{D}} e^{-\frac{2 L}{D \tau}} L^{\frac{2 \gamma
   }{D}+1}}{\Gamma \left(\frac{2 (D+\gamma )}{D}\right)} & L \le L_f \\
 \frac{4^{\frac{\gamma +D}{D}} (D \tau)^{-\frac{2 (\gamma +D)}{D}} e^{-\frac{2 L}{D \tau}} L_f^{\frac{2 \gamma
   }{D}+1}}{\Gamma \left(\frac{2 (D+\gamma )}{D}\right)} & L \ge L_f
   \ .
\end{cases}
\end{align}
Note, that we additionally transformed $L_f$ using the transformation from $x$ to $y$. We validate Eq.~\eqref{sol_bessel_mult} by substitution into the to Eq.~\eqref{sup_tr_lv_bessel_tsa_simp_mult} corresponding reverse time Fokker-Planck for $L>L_f$ and $L<L_f$, respectively. To check whether Eq.~\eqref{sol_bessel_mult} decays with a rate equal to the forward hitting time, we calculated it from the cumulative of  Eq.~\eqref{sol_bessel_mult} and compare it to the hitting time distribution 
\begin{align}
\label{sup_hittingTdist_transf_fwd_Bessel}
\rho_0(\tau|L_f)
=
 \frac{L_f 4^{\frac{\gamma }{D}} e^{-\frac{2 L_f}{D \tau}} \left(\frac{D
   \tau}{L_f}\right)^{-\frac{2 \gamma }{D}}}{\gamma  \tau^2 \Gamma \left(\frac{2 \gamma }{D}\right)}
\end{align}
of the Ito transformed forward Bessel process stated in Eq.~\eqref{sup_fwd_sde_bessel_mult_ForceAndNoise}. Formally, this hitting time distribution is obtained starting from the known solution of the forward Bessel process \cite{bray2000random,edgar2011bessel} (reproduced with our notation in \cite{lenner2023reversetime}): We apply the same transformations and substitutions as for Eq.~\eqref{sol_bessel_mult}, and the definition of the hitting time distribution Eq.~\eqref{eq:def_hit_dist}, to obtain the final result stated in Eq.~\eqref{sup_hittingTdist_transf_fwd_Bessel}.

To confirm the validity of both the reverse time TSA-SDE with multiplicative noise and the solution of aboves TSA-dynamics, we calculate the mean 
\begin{align}
\overline{L}(\tau) =
 \frac{D^2 \tau^2 \left(\Gamma \left(\frac{2 \gamma }{D}+3\right)-\Gamma \left(\frac{2 \gamma }{D}+3,\frac{2
   L_f}{D \tau}\right)\right)+L_f 2^{\frac{2 \gamma }{D}+1} e^{-\frac{2 L_f}{D \tau}} (D \tau+2
   L_f) \left(\frac{D \tau}{L_f}\right)^{-\frac{2 \gamma }{D}}}{2 \tau (2 \gamma +D) \left(\Gamma
   \left(\frac{2 \gamma }{D}+1\right)-\Gamma \left(\frac{2 \gamma }{D}+1,\frac{2 L_f}{D
   \tau}\right)\right)}
\end{align}
and variance
\begin{align}
 \sigma_L^2(\tau) 
 =
 \frac{
 D^3 \tau^3 \left(\!\Gamma \! \left(\frac{2 \gamma }{D}+4 \!\right)-\Gamma \left(\frac{2 \gamma }{D}+4,\frac{2
   L_f}{D \tau}\!\right)\!\right)+L_f 4^{\frac{\gamma +D}{D}} e^{-\frac{2 L_f}{D \tau}} \left(\! D^2
   \tau^2+2 D \tau L_f+2 L_f^2 \!\right) 
   \left(\!\frac{D \tau}{L_f}\!\right)^{-\frac{2 \gamma }{D}}
   }{
   4 \tau (2
   \gamma +D) \left(\!\Gamma \left(\frac{2 \gamma }{D}+1\right)-\Gamma \left(\!\frac{2 \gamma }{D}+1,\frac{2
   L_f}{D \tau}\!\right)\!\right)}
 - \overline{L}(\tau)^2
\end{align}
with respect to Eq.~\eqref{sol_bessel_mult} after normalization. In Fig.~3 of the main text, we compare this mean and variance to the mean and variance of the forward dynamics after target state alignment. 
Given sufficiently many forward trajectories for the mean and variance to converge, we find perfect agreement.

In general, care must be taken when using Ito's lemma to construct equivalent reverse time TSA dynamics from known expressions. First, the hitting time distribution of the forward problem must be recalculated with respect to the transformed forward transition probability $P^\mathrm{fw}(\widehat{L},t)$. And second, using arbitrary initial distributions $P^\mathrm{in}(L)$, it is important to clarify in which variable space the initial distribution is defined and whether we need to transform it to the new coordinates. 


\subsection{TSA-dynamics close to the target state}
The dynamics both spatially and temporally close to the target state can be deduced from Eqs.~\eqref{sup_tr_lv_tsa_mult},\eqref{sup_freeE_force_multNoise},\eqref{sup_H_F_dep_on_in_cond} and \eqref{sup_killing_measure_def}. We here briefly summarize the argument we previously gave in Lenner et.~al.~\cite{lenner2023reversetime} for the analogous case of state independent noise. Assuming well separated initial and final conditions, the dependency on the initial conditions in the Langevin equation Eq.~\eqref{sup_tr_lv_tsa_mult} can be neglected. We set the initial condition dependent term $H(L)$, defined in Eq.~\eqref{sup_H_F_dep_on_in_cond}, to one and the killing measure $k(L,\tau)$, defined in Eq.~\eqref{sup_killing_measure_def} to zero. We arrive at
\begin{align}
\label{sup_tr_lv_tsa_CFB_mult}
  dL(\tau) =
\left(
      f(L) 
      +D(L) 
      \  
      \frac{\partial}{\partial 
L}
 \log\left(
  \int_{L_\mathrm{ts}}^{L} dL' \;
   e^{-
  \int^{L'}
 \frac{2 f(L'')}{D(L'')} dL''
 } 
 \right)  
 \right)
       d\tau
      +
      \sqrt{D(L)} \ d W_\tau
     \ .
\end{align}
The free energy force Eq.~\eqref{sup_freeE_force_multNoise} is now independent of any initial conditions of the forward process and reads
\begin{align}
\label{freeEforce_multN_closeToCompl}
f^\mathcal{F}(L)
=
       D(L) 
      \  
      \frac{\partial}{\partial 
L}
 \log\left(
  \int_{L_\mathrm{ts}}^{L} dL' \;
   e^{-
  \int^{L'}
 \frac{2 f(L'')}{D(L'')} dL''
 } 
 \right)
 \ .
\end{align}
To form a better understanding of how these TSA-dynamics differ from their forward counterpart, we study the case of power law like dynamics both in the force 
\begin{align}
 f(L) = -\gamma L^\alpha
\end{align}
and diffusion term
\begin{align}
 D(L)=D L^\beta
\end{align}
with $D>0$ and $\alpha , \beta, \gamma \in \mathbb{R}$. We observe, that, when evaluating the fraction $-\frac{2 f(L)}{D(L)}:=\frac{2\gamma L^{\alpha-\beta}}{D}$ in Eq.~\eqref{freeEforce_multN_closeToCompl}, the exponential simplifies to an expression formally indistinguishable from the constant $D$ case stated in ~\cite{lenner2023reversetime}, where we used $-\frac{2 f(L)}{D}:=\frac{2\gamma L^{\alpha}}{D}$. The $\log$-part of the free energy force for multiplicative noise (Eq.~\eqref{freeEforce_multN_closeToCompl}) is therefore functionally identical to our previous result with constant noise strength and $\alpha \to \alpha-\beta $. The free energy force for multiplicative noise can therefore be obtained from our previous result using the mapping
\begin{align}
\label{freeE_subst_formula}
f^\mathcal{F}(L)=L^\beta f_0^\mathcal{F}(L,\alpha-\beta) 
\ .
\end{align}
where $f_0^\mathcal{F}(L,\alpha)$ denotes the free energy force with constant noise. The term $L^\beta$ corrects for the now explicitly $L$ dependent diffusion constant $D(L)$ in front of the logarithm of the free energy force (Eq.~\eqref{freeEforce_multN_closeToCompl}). Stated explicitly we find
\begin{align}
  \label{sup_freeEforce_fwd_powerlaw_mult}
f^\mathcal{F}(L)
=
L^\beta
 \frac{D(\alpha-\beta+1) \left(-\frac{2 \gamma }{D(\alpha-\beta+1)}\right)^{\frac{1}{\alpha-\beta+1}} e^{\frac{2 \gamma  L^{\alpha-\beta+1}}{D(\alpha-\beta+1)}}}{\Theta(\alpha-\beta+1)  \Gamma \left(\frac{1}{\alpha-\beta+1}\right)-\Gamma
   \left(\frac{1}{\alpha-\beta+1},-\frac{2 L^{\alpha-\beta+1} \gamma }{D(\alpha-\beta+1)}\right)}
   \qquad \alpha -\beta \neq -1
   \ .
\end{align}
The case $\alpha-\beta=1$ can be obtained using the same mapping as above, or simply by evaluating Eq.~\eqref{sup_tr_lv_tsa_CFB_mult}. We find
\begin{align}
f^\mathcal{F}(L)
    &= 
    L^\beta
 \frac{D(\alpha-\beta+1) \left(-\frac{2 \gamma }{D(\alpha-\beta+1)}\right)^{\frac{1}{\alpha-\beta+1}} e^{\frac{2 \gamma  L^{\alpha-\beta+1}}{D(\alpha-\beta+1)}}}{\Theta(\alpha-\beta+1)  \Gamma \left(\frac{1}{\alpha-\beta+1}\right)-\Gamma
   \left(\frac{1}{\alpha-\beta+1},-\frac{2 L^{\alpha-\beta+1} \gamma }{D(\alpha-\beta+1)}\right)}
    \qquad 
    &\mathrm{valid \ for} \qquad \alpha -\beta \neq -1
\intertext{and}
\label{sup_freeE_force_alpha_simple_diag}
f^\mathcal{F}(L)
    &= (2\gamma+D) L^\alpha
        \qquad 
    &\mathrm{valid \ for} \qquad \alpha -\beta = -1
    \ .
\end{align}
$\Gamma(z)$ is the gamma-function, $\Gamma \left( \nu,z \right)$ the upper incomplete gamma function and $\Theta(\nu)$ the Heaviside step function.

%


\subsection{Exactly solvable TSA dynamics close to the target state}
To study the quality of the TSA approximation for multiplicative noise, we 
exclusively consider the case $\alpha -\beta = -1$. For this case the TSA dynamics close to the target state are most simple. Using the free energy force as defined in Eq.~\eqref{sup_freeE_force_alpha_simple_diag}, the reverse time Langevin equation Eq.~\eqref{sup_tr_lv_tsa_mult} become simple, i.e.
\begin{align}
  \label{sup_freeEforce_fwd_powerlaw_mult_minone}
dL(\tau)
=
(\gamma+D) L^\alpha
        d\tau
      +
      \sqrt{D L^{\alpha+1}} \ d W_\tau
      \qquad \alpha -\beta = -1
     \ .
\end{align}
with $\alpha<1$. 
For these values of $\alpha$ we can exploit Ito's lemma \cite{gardiner1985handbook} and transform 
the reverse time TSA version of the Bessel process \cite{bray2000random} with well separated initial and final conditions 
\begin{align}
\label{sup_tr_lv_ness_bessel_force}
dL(\tau)
=
\frac{\gamma +D}{L}
 d\tau
   + \sqrt{D} \; dW_\tau
   \qquad \alpha =  -1
   \ .
\end{align}
(here restated from Lenner et.~al.~\cite{lenner2023reversetime}) 
into Eq.~\eqref{sup_freeEforce_fwd_powerlaw_mult_minone}. We use the substitution $L_\mathrm{new}=L_\mathrm{old}^\frac{2}{1-\alpha}$, and replace
 $D_\mathrm{old}=\frac{(1-\alpha)^2}{4}D_\mathrm{new}$ and $\gamma_\mathrm{old}=\frac{1-\alpha}{2} \gamma_\mathrm{new}+ \frac{D_\mathrm{new}}{8}(1-\alpha^2)$. 
 Note that for $\alpha>1$ the used target state at $L_\mathrm{ts}$ would change its position from zero to infinity under the given transformation. With a different spirit, this reiterates our result discussed in the main text, that dynamics, with $\alpha -\beta = -1$ and $\alpha\ge1$, can not reach the target state in finite time. 

In the next step we use this transformation starting from the known solution for Eq.~\eqref{sup_tr_lv_ness_bessel_force} stated in Lenner et.~al.~\cite{lenner2023reversetime} and originally obtained by \cite{bray2000random}, to solve  Eq.~\eqref{sup_freeEforce_fwd_powerlaw_mult_minone} for $\alpha<1$ and with target state $L_\mathrm{ts}=0$. We follow the same logic as discussed in sub-section \ref{sec_Exactly solvable reverse time TSA dynamics with multiplicative noise}. For the density we find
\begin{align}
R(L,\tau)
=
 -\frac{(\alpha -1) 2^{\frac{2 \gamma -\alpha  D+2 D}{D-\alpha  D}} \left((\alpha -1)^2 D \tau \right)^{\frac{2
   \gamma +D}{(\alpha -1) D}-1} L^{-\alpha +\frac{2 \gamma }{D}+1} e^{-\frac{2 L^{1-\alpha }}{(\alpha -1)^2 D
   \tau }}}{\Gamma \left(\frac{-\alpha  D+2 D+2 \gamma }{D-D \alpha }\right)}
 \qquad \alpha <1 \ ,
\end{align}
leaving us with well defined  mean
\begin{align}
\label{mean_gen_bessel}
\overline{L}(\tau)
=
 \frac{2^{\frac{1}{\alpha -1}} (1-\alpha )^{-\frac{2}{\alpha -1}}  \Gamma
   \left(\frac{-\alpha  D+3 D+2 \gamma }{D-D \alpha }\right)}{\Gamma \left(\frac{-\alpha  D+2 D+2 \gamma }{D-D
   \alpha }\right)}
   (D \tau)^{\frac{1}{1-\alpha }}
   \qquad \alpha <1 \ ,
\end{align}
and variance
\begin{align}
\label{var_gen_bessel}
\sigma^2_L(\tau)
=
\frac{4^{\frac{1}{\alpha -1}} (1-\alpha )^{-\frac{4}{\alpha -1}}  \left(\Gamma
   \left(\frac{-\alpha  D+2 D+2 \gamma }{D-D \alpha }\right) \Gamma \left(\frac{-\alpha  D+4 D+2 \gamma }{D-D
   \alpha }\right)-\Gamma \left(\frac{-\alpha  D+3 D+2 \gamma }{D-D \alpha }\right)^2\right)}{\Gamma
   \left(\frac{-\alpha  D+2 D+2 \gamma }{D-D \alpha }\right)^2}
   (D \tau)^{-\frac{2}{\alpha -1}}
\end{align}
for $\alpha <1$. Despite the complicated prefactors, the dependency on $\tau$ is simple.

\subsection{The small $L$ expansion for TSA dynamics close to the target state} 
\label{The_small_L_expansion_for_multiplicative_noise_TSA_dynamics_close_to_the_target_state}
We next consider the expansion of the free energy force in small orders of $L$.
The expansion of the free energy force $f_0^\mathcal{F}(L)$ with constant noise variance $D$, is derived in Lenner et.~al.~\cite{lenner2023reversetime}. Using the substitution formula Eq.~\eqref{freeE_subst_formula}, together with our previous results for the constant noise variance, allows us to map the constant noise variance case to the multiplicative noise case. We directly arrive at the low order expressions for the 
noise driven ($\alpha \ge \beta$)
\begin{align}
\label{sup_tr_lv_tsa_sn_multN_alpha_ge_beta_simp}
  dL(\tau) =
\left(
      D L^{\beta-1}
      -
      \gamma \frac{\alpha-\beta}{\alpha-\beta+2} L^{\alpha}
 \right)
       d\tau
      +
      \sqrt{D \, L^\beta} \ d W_\tau
\end{align}
 and force driven ($\alpha < \beta$)
 \begin{align}
\label{sup_tr_lv_tsa_sn_multN_alpha_sm_beta_simp}
  dL(\tau) =
  \left(
    \gamma L^\alpha
      -  D (\alpha-\beta) L^{\beta-1} 
 \right)
       d\tau
      +
      \sqrt{D \, L^\beta} \ d W_\tau
\end{align}
case. Both the noise driven and force driven case are valid in the limit of small $L$. In the force driven case we additionally require $D$ to be small. See  Lenner et.~al.~\cite{lenner2023reversetime} for the mathematical details.

\subsection{Moments for noise driven TSA dynamics}
\label{Moments_for_the_noise_driven_TSA_dynamics_with_multiplicative_noise}
For $\alpha \ge \beta$, the dynamics close to the target state are noise dominated, as we read off from Eq.~\eqref{sup_tr_lv_tsa_sn_multN_alpha_ge_beta_simp}. To lowest order this expression simplifies to
\begin{align}
\label{langevin_tsa_pure_multN}
  dL(\tau) =
      D L^{\beta-1}
       d\tau
      +
      \sqrt{D \, L^\beta} \ d W_\tau
      \qquad
      \mathrm{for}
      \qquad
      \alpha \ge \beta
     \ ,
\end{align}
which is analytically tractable.
The solution of Eq.~\eqref{langevin_tsa_pure_multN} with $\beta<2$ is a special case of 
Eq.~\eqref{sup_freeEforce_fwd_powerlaw_mult_minone}. We can therefore re-use our previous results for mean and variance stated in Eq.~\eqref{mean_gen_bessel} and Eq.~\eqref{var_gen_bessel}.
With $\gamma=0$ and $\alpha=\beta-1$, we find for mean
\begin{align}
\label{MultNoiseDriven_mean}
\overline{L}(\tau)
=
 \frac{2^{\frac{\beta -1}{\beta -2}} (2-\beta )^{-\frac{2}{\beta -2}} \Gamma \left(-\frac{2}{\beta -2}\right) }{\Gamma \left(\frac{1}{2-\beta }\right)}
 (D\tau)^{\frac{1}{2-\beta }}
\end{align}
and 
variance
\begin{align}
\label{MultNoiseDriven_var}
\sigma_L^2(\tau)
=
\frac{(2-\beta )^{-\frac{4}{\beta -2}} \left(3\ 4^{\frac{1}{\beta -2}} \Gamma \left(\frac{1}{2-\beta }\right)
   \Gamma \left(-\frac{3}{\beta -2}\right)-4^{\frac{\beta -1}{\beta -2}} \Gamma \left(-\frac{2}{\beta
   -2}\right)^2\right) }{\Gamma \left(\frac{1}{2-\beta }\right)^2}
   (D \tau)^{\frac{1}{1-\beta/2}}
   \ .
\end{align}
Despite their lengthy prefactors both mean and variance show a very simple dependency on $\tau$.
For $\beta=0$, both terms simplify to the case of constant $D$, with a square-root dependence of the mean and a linear dependence on $\tau$ for the variance.


\subsection{Weak noise approximation of moments for force driven TSA ensembles}
The derivation of the small noise moments given multiplicative noise closely follows the special case with constant $D$ discussed in Lenner et.~al.~\cite{lenner2023reversetime}. We follow the approach discussed in Gardiner \cite{gardiner1985handbook}.

We start our derivation with a rewritten version of Eq.~\eqref{sup_tr_lv_tsa_sn_multN_alpha_sm_beta_simp}
\begin{align}
  dL = a(L) d\tau + \epsilon^2 b(L) d\tau + \epsilon \; c(L)  d W_\tau
  \ ,
\end{align}
with  $\sqrt{D}$ substituted by the order parameter $\epsilon$. We next expand 
\begin{align}
\label{sup_epsilon_expansion}
 L(\tau) = L_0(\tau) + \epsilon L_1(\tau) + \epsilon^2 L_2(\tau) + ...
\end{align}
for small $\epsilon$ and around the deterministic solution $L_0(\tau)$. Similarly we expand $a(L)$, $b(c)$ and $c(L)$. Assuming that $a(L)$ can be written as
\begin{align}
 a(L) = a(L_0 + \epsilon L_1 + \epsilon^2 L_2 + ...)
      = a_0(L_0) + \epsilon a_1(L_0,L_1) + \epsilon^2 a_2(L_0,L_1,L_2)
      + ... \ ,
\end{align}
a general expansion reads
\begin{align}
 a(L) = a\left(L_0 + \sum_{m=1}^\infty \epsilon^m L_m \right)
 = \sum_{p=0}^\infty \frac{1}{p!} \frac{d^p a(L_0)}{d L_0} 
 \left(
 \sum_{p=0}^\infty \epsilon^m L_m
 \right)^p
 \ .
\end{align}
An analogous expression holds for $b(L)$ and $c(L)$.
After sorting terms we find for the first three terms
\begin{align}
 &a_0(L_0) = a(L_0) 
 \\
 &a_1(L_0,L_1) = L_1 \frac{d a(L_0)}{d L_0}
 \\
 \label{sup_a2}
 &a_2(L_0,L_1,L_2) = L_2 \frac{d a(L_0)}{d L_0} 
 + \frac{1}{2} L_1^2 \frac{d^2 a(L_0)}{d L_0^2}
 \ .
\end{align}
Following the same expansion scheme for $b(L)$ and $c(L)$ we arrive at an ordered set of SDEs
\begin{align}
\label{sup_dx0_mult}
 &d L_0 = a(L_0) d\tau
 \\
\label{sup_dx1_mult}
 &d L_1 = a_1(L_1,L_0) d\tau + c(L_0) d W_\tau
 \\
 \label{sup_dx2_mult}
 &d L_2 = a_2(L_2,L_1,L_0) d\tau + b(L_0) d\tau + c(L_1,L_0) dW_\tau
 \ ,
\end{align}
which we truncate after the second contributing order to the full solution $L(\tau)$. For $f(L)=-\gamma L^\alpha$ and $D(L)=D L^\beta$, as defined above, the terms evaluate as follows. The zeroth order term 
\begin{align}
\label{sup_sn_Lzero}
 dL_0 = \gamma L_0^\alpha d\tau
 \ .
\end{align}
defines the solution of the deterministic dynamics. With the target state $L_\mathrm{ts}=0$ at the boundary, the solution is given as
\begin{equation}
\label{sup_x0_res}
L_0 = \left( (1-\alpha) \gamma \tau \right)^{\frac{1}{1-\alpha}} 
\qquad \mathrm{with} \ \alpha <1
\ .
\end{equation}
The equation for the first correction to the deterministic solution Eq.~\eqref{sup_dx1_mult} reads 
\begin{align}
\label{sup_dx1_expl_mult}
 d L_1 = L_1 k\left(L_0(\tau)\right) d\tau + c(L_0(\tau)) d W_\tau
 \ .
\end{align}
The time dependent drift coefficient evaluates to
\begin{align}
\label{sup_k}
  k\left(L_0\right) = \frac{d a(L_0)}{d L_0} = \gamma \frac{d L_0^\alpha}{d L_0}
	= \gamma \alpha L_0^{\alpha-1}
	= \gamma \alpha \left( (1-\alpha) \gamma \tau \right)^{-1}
\ ,
\end{align}
and the coefficient of the diffusion term is given as
\begin{align}
\label{sup_c_mult}
 c(L_0(\tau) )
 = 
 L_0^\frac{\beta}{2} 
 = 
\left( (1-\alpha) \gamma \tau \right)^{\frac{\beta}{2-2\alpha}} 
\ ,
\end{align}
where for both we used the explicit solution for $L_0$ stated in Eq.~\eqref{sup_x0_res}.
The formal solution to Eq.~\eqref{sup_dx1_expl_mult} then reads
\begin{align}
\label{sup_x1_res_mult}
 L_1(\tau) 
 &= \int_0^\tau  
 c(L_0(\tau'))
 e^{
 \int_{\tau'}^\tau k(L_0(s)) ds
 }
 d W_{\tau'}
 \ ,
\intertext{or explicitly written}
 &= 
 \int_0^\tau
 \left( (1-\alpha) \gamma \tau' \right)^{\frac{\beta}{2-2\alpha}} 
 \left(\frac{\tau}{\tau'}\right)^{-\frac{\alpha
   }{\alpha -1}}
   d W_{\tau'} 
   \ ,
\end{align}
 with $k(L_0(s))$ taken from Eq.~\eqref{sup_k} and $c(L_0(\tau'))$ from Eq.~\eqref{sup_c_mult}. As above, we assume the target state at $L(0)=0$ and thus $L_1(0)=0$.

To formally find an analytic expression for the second order contribution $L_2(\tau)$ we must solve Eq.~\eqref{sup_dx2_mult}. However, as we will see below, to construct expressions for mean, variance and two-time covariance up to order $D$ it is sufficient to determine the ensemble average of $L_2(\tau)$. Averaging over Eq.~\eqref{sup_dx2_mult}, we arrive at the ordinary differential equation (ODE)
\begin{equation}
\label{eq:l2_ode}
 \frac{d \langle L_2 \rangle}{d\tau}
 =
 \frac{d a(L_0)}{dL_0} \langle L_2 \rangle
 + \frac{1}{2} \frac{d^2 a(L_0)}{d L_0^2} \langle L_1^2 \rangle
 + b(L_0)
 \ ,
\end{equation}
where $b(L_0)$ is given as
\begin{align}
 b(L_0) = - (\alpha-\beta) L_0^{\beta-1}
 .
\end{align}
We next exploit the expressions for $L_0$, $L_1$ and $\langle L_2 \rangle$ to construct moments for the homogeneous dynamics. We then use the obtained expressions to generalized the results to heterogeneous force laws.

For the ensemble mean we substitute $L$ with its expansion up to order $D$ to obtain
\begin{align}
\label{sup_def_sn_mean}
\langle L(\tau) \rangle 
&= 
\langle L_0(\tau) \rangle 
+ \sqrt{D} \langle L_1(\tau) \rangle 
+ D \langle L_2(\tau) \rangle
+ \mathcal{O}(D^\frac{3}{2}) \ .
\intertext{The zero-th order term is simply the deterministic solution. The next leading order term 
$\langle L_1(\tau) \rangle$ evaluates to zero, as the Wiener increment denotes a zero mean white noise stochastic process. The mean up to order $D$ is thus constituted as
}
&= 
\langle L_0(\tau) \rangle 
+ D \langle L_2(\tau) \rangle
+ \mathcal{O}(D^\frac{3}{2}) \ .
\end{align}
For the variance 
\begin{align}
\label{sup_def_sn_var}
\sigma_L^2(\tau)
&=
\langle L(\tau)^2 \rangle - \langle L(\tau) \rangle^2
=
\langle \left(L_0(\tau) + \sqrt{D}L_1(\tau) + D L_2(\tau) \right)^2 \rangle 
- 
\langle L_0(\tau) + \sqrt{D}L_1(\tau) + D L_2(\tau) \rangle^2
+ \mathcal{O}(D^\frac{3}{2})
\intertext{
only one term up to order $D$ survives. 
The zero-th order term is a deterministic expression and thus evaluates to zero. Furthermore, the $L_0$ $L_1$ cross terms evaluate to zero with $L_0$ deterministic and 
$\langle L_1(\tau) \rangle =0$. Analogously with $L_0$ deterministic the cross $L_0$ $L_2$ term cancels. The only surviving term is thus the order $D$ term
}
&=
D \langle L_1^2(\tau) \rangle + \mathcal{O}(D^\frac{3}{2})
\end{align}
For the two-time covariance the same observations as for the variance hold. Only the doubly $L_1$ dependent term
\begin{align}
  C(\tau,\tau') &= 
\langle
\left(
L(\tau) - \langle L(\tau) \rangle
\right)
\left(
L(\tau') - \langle L(\tau') \rangle 
\right)
\rangle
=
D \langle L_1(\tau) L_1(s) \rangle + \mathcal{O}(D^\frac{3}{2})
\end{align}
survives. 
We next evaluate all contributing expressions for mean, variance and two-time covariance. The combined expressions are stated in the main text. For the sole contribution to the variance we find
\begin{equation}
\label{sup_var_epsilon_expansion_mult}
 \langle L_1^2(\tau) \rangle 
 =  
\int_0^\tau
 \left( (1-\alpha) \gamma \tau' \right)^{\frac{\beta}{1-\alpha}} 
 \left(\frac{\tau}{\tau'}\right)^{-\frac{2 \alpha
   }{\alpha -1}}
   d \tau'
   =
   \frac{D}{\gamma}
   \frac{ ((1 -\alpha)\gamma \tau)^{\frac{1-\alpha+\beta }{1-\alpha }}}{1-3 \alpha
   +\beta}
   \qquad \mathrm{for} \ \alpha <0
   \ .
\end{equation}
Similar to the variance, the two time covariance is built from the product of two $L_1(\tau)$ terms. Evaluating these terms we find
\begin{align}
\label{sup_2time_cov_epsilon_expansion_mult}
 \langle L_1(\tau) L_1(\tau') \rangle
 &=
\int_0^{\mathrm{min}(\tau,\tau')} 
 \left( (1-\alpha) \gamma s \right)^{\frac{\beta}{1-\alpha}} 
  \left(\frac{\tau}{s}\right)^{-\frac{ \alpha
   }{\alpha -1}}
     \left(\frac{\tau'}{s}\right)^{-\frac{ \alpha
   }{\alpha -1}}
   d s
\notag
   \\
   &=
\begin{cases}
  \frac{((1-\alpha ) \gamma  \tau')^{-\frac{\alpha }{\alpha -1}} ((1-\alpha ) 
   \gamma \tau)^{\frac{1 -2 \alpha +\beta}{1-\alpha }}}{\gamma \left(1 -3 \alpha +\beta \right)}
   &\qquad \mathrm{for} \ \tau < \tau'
   \\
  \frac{((1-\alpha ) \gamma \tau)^{-\frac{\alpha }{\alpha -1}} ((1-\alpha )  
   \gamma \tau')^{\frac{1 -2 \alpha +\beta}{1-\alpha }}}{\gamma \left(1 -3 \alpha +\beta \right)}
   &\qquad \mathrm{for} \ \tau > \tau' 
   \\
    \frac{( (1 -\alpha) \gamma \tau)^{\frac{1 -\alpha +\beta}{1-\alpha }}}{
 \gamma \left(1 - 3 \alpha + \beta \right) }
   &\qquad \mathrm{for} \ \tau = \tau'
   \ .
\end{cases}
\qquad \mathrm{and} \ \alpha <0
\end{align}
In the last step we solve the ODE Eq.\eqref{eq:l2_ode} to find the order $D$ contribution to the mean 
\begin{equation}
\label{eq:sol_l2}
\langle L_2(\tau) \rangle
=
\frac{\left(7 \alpha ^2-\alpha  (8 \beta +3)+2 \beta  (\beta +1)\right) ((1-\alpha) \gamma 
   \tau)^{\frac{\alpha -\beta }{\alpha -1}}}{2 \gamma  (2 \alpha -\beta ) (3 \alpha -\beta
   -1)}
\qquad \mathrm{for} \ \alpha <0
\ .
\end{equation}
Collecting terms of different orders we arrive at the small noise expressions for mean, variance and covariance stated in the main text.

\section{The Feller boundary classification scheme}
In this study we establish a general understanding of stochastic dynamics close to target states, which follows the forward-time Langevin equation 
\begin{align}
 d\widehat{L} = f(\widehat{L}) dt +\sqrt{D(\widehat{L})} dW_t .
\end{align}
Based on the assumption that both the deterministic advective force $f(\widehat{L})$ and the diffusion 'constant' $D(\widehat{L})$ are continuous functions in $\widehat{L}$, their relevant behavior close to the target state can be well approximated by its lowest order Taylor expansion, i.e. by a power law. Explicitly, this means $f(\widehat{L})=-\gamma \widehat{L}^\alpha$ and $D(\widehat{L})=\widehat{L}^\beta$ close to a target state at $\widehat{L}_\mathrm{ts}=0$. For general exponents the question arises whether the target state can be reached in the first place and if yes "how" it is reached. This means we would like to obtain a classification of the boundaries in dependence of the exponents and visualize the results in $\alpha-\beta-$phase-space. The Feller boundary classification scheme provides the tools to do this analytically and to obtain a phase diagram which is almost impossible to come by numerically. To our knowledge such a general characterization based on exponents $\alpha$ and $\beta$ has not been given and only a few very specific choices of $\alpha$ and $\beta$ are discussed in the literature \cite{edgar2011bessel,trabelsi2016boundary, albanese2007transformations}. In this section we briefly recapitulate Feller's boundary classification scheme -- which is typically stated in the for many physicist unfamiliar martingal formalism -- in a form better recognizable for many physicists. 

\subsection{Intuitive explanation}
Feller's boundary classification is based on two questions. Given that the dynamics lives in an interval 
$(\widehat{L_a},\widehat{L_b})$, (i) is there a finite probability $\mathbb{P}_{\widehat{L}}(T_a<T_b)$ to pass through one of the boundary (at time $T_a$) before exiting through the other (at time $T_b$)? (ii) Does this event occur on average in finite time, that is, what is the mean first passage time $\mathbb{E}_{\widehat{L}}(T)$? That only these two questions are answered by Feller's boundary classification is sometimes obscured by the fact that the classification scheme is based on four criteria which are derived from $\mathbb{P}_{\widehat{L}}(T_a<T_b)$ and $\mathbb{E}_{\widehat{L}}(T)$. Feller's scheme differentiates four boundary classes which are given by those 4 criteria. Assuming that the left endpoint of the interval approaches the target state $\widehat{L_a} \rightarrow \widehat{L_{ts}}$, the criteria are given by the expressions $S[\widehat{L}_\mathrm{ts},\widehat{L}_b]$, $M[\widehat{L}_\mathrm{ts},\widehat{L}_b]$, $\Sigma[\widehat{L}_\mathrm{ts},\widehat{L}_b]$ and $N[\widehat{L}_\mathrm{ts},\widehat{L}_b]$, which will be defined below in the following section. Here, we are only giving some intuition:  
\begin{itemize}
\item $S[\widehat{L}_\mathrm{ts},\widehat{L}_b]$ and $M[\widehat{L}_\mathrm{ts},\widehat{L}_b]$ map arbitrary one-dimensional stochastic dynamics on the simple Brownian motion case 
\begin{itemize}
\item $S[\widehat{L}_\mathrm{ts},\widehat{L}_b]$ does the transformation in space and is therefore called scale function. In addition, it constitutes a measure of the time for the dynamics to reach the target state $\widehat{L}_\mathrm{ts}$ before reaching the second boundary
\item $M[\widehat{L}_\mathrm{ts},\widehat{L}_b]$ does the transformation in time  and is therefore called speed measure. It measures the speed of the dynamics in the vicinity of the target state $\widehat{L}_\mathrm{ts}$ 
\end{itemize}
\item $\Sigma[\widehat{L}_\mathrm{ts},\widehat{L}_b]$ quantifies the average time to reach the boundary when starting from an arbitrary state $\widehat{L}$ within the interval 
\item $N[\widehat{L}_\mathrm{ts},\widehat{L}_b]$ quantifies the average time to reach a state $\widehat{L}$ within the interval when starting from the target state $\widehat{L_{ts}}$
\end{itemize}
The transformation to the Brownian motion case makes the asymptotic behavior of those criteria analytically accessible even if it is not possible to find an explicit expression for the transition density. Their asymptotic behavior close to the target state then determines the following boundary classes:
\begin{itemize}
 \item regular: The boundary can be reached in finite time and points in the interval are accessible if starting from the boundary. 
 $S[\widehat{L}_\mathrm{ts},\widehat{L}_b]<\infty$, $M[\widehat{L}_\mathrm{ts},\widehat{L}_b]<\infty$, $\Sigma[\widehat{L}_\mathrm{ts},\widehat{L}_b]<\infty$ and $N[\widehat{L}_\mathrm{ts},\widehat{L}_b]<\infty$
 \item exit: The boundary can be reached in finite time but entering the definition interval from the boundary is not possible.  $S[\widehat{L}_\mathrm{ts},\widehat{L}_b]<\infty$, $M[\widehat{L}_\mathrm{ts},\widehat{L}_b]=\infty$, $\Sigma[\widehat{L}_\mathrm{ts},\widehat{L}_b]<\infty$ and $N[\widehat{L}_\mathrm{ts},\widehat{L}_b]=\infty$
 \item entrance: The boundary can not be reached from the inside of the interval but starting at the boundary leads to entrance into the interval in finite time.  $S[\widehat{L}_\mathrm{ts},\widehat{L}_b]=\infty$, $M[\widehat{L}_\mathrm{ts},\widehat{L}_b]<\infty$, $\Sigma[\widehat{L}_\mathrm{ts},\widehat{L}_b]=\infty$ and $N[\widehat{L}_\mathrm{ts},\widehat{L}_b]<\infty$
 \item natural: Natural boundary conditions summarize all cases where the boundary can neither be reached or exited in finite time, that is $\Sigma[\widehat{L}_\mathrm{ts},\widehat{L}_b]=\infty$ and $N[\widehat{L}_\mathrm{ts},\widehat{L}_b]=\infty$ and $S[\widehat{L}_\mathrm{ts},\widehat{L}_b]$, $M[\widehat{L}_\mathrm{ts},\widehat{L}_b]$ arbitrary.
\end{itemize}

\subsection{Derivation of the criteria}
Because the four criteria are not easily understood by their mere definitions, we are giving a brief derivation. It is highly recommended to study Feller's derivation in more detail and exactness \cite{feller1951diffusion, feller1952parabolic} and reviewed in \cite{ito2012diffusion, taylor1981second}. In addition, we are partially following a nicely readable review \cite{trabelsi2016boundary} and a lecture script \cite{etheridge2021stochasticscript}.  \\ 

Both the first passage probability and the mean first passage time are accessible via the first passage time distribution
\begin{align}
\label{eq:hittingTimeDist}
 \rho(T|\widehat{L})= \frac{\partial}{\partial T} 
 \left(
 1- \int_{\widehat{L}_a}^{\widehat{L}_b} d\widehat{L}_f \; P(\widehat{L}_f,T|\widehat{L},0)
 \right)
 \ .
\end{align}
The first passage probability 
\begin{align}
 u(\widehat{L}) 
 &=
 \int_0^\infty \rho(T|\widehat{L}) \; dT 
\quad \mathrm{with} \quad u(\widehat{L}_a)=1 \ , \ u(\widehat{L}_b)=0
 \notag \\
 &=
 \mathbb{P}_{\widehat{L}}(T_a<T_b)  
\end{align}
is obtained by integration of its density, the mean first passage time
\begin{align}
 v(\widehat{L})
  &=
 \int_0^\infty T \rho(T|\widehat{L}) \; dT \quad \mathrm{with} \quad v(\widehat{L}_a)=0 \ , \ v(\widehat{L}_b)=0
 \notag \\
 &=
 \mathbb{E}_{\widehat{L}}(T)
\end{align}
by averaging. Exact expressions for the hitting time distribution can be obtained from Eq.~\eqref{eq:hittingTimeDist}, and thus from the solution of the backward/forward FPE. Alternatively, $\rho(T|\widehat{L})$ can be obtained directly from its defining backward FPE equation  \cite{gardiner1985handbook}
\begin{align}
\label{eq:bwFPEhittingTimeDist}
 \frac{\partial}{\partial T} \rho(T|\widehat{L})
 =
 -
 f(\widehat{L})
 \frac{\partial}{\partial \widehat{L}} \rho(T|\widehat{L})
 -
 \frac{D(\widehat{L})}{2}
 \frac{\partial^2}{\partial \widehat{L}^2}
 \rho(T|\widehat{L})
 \ ,
\end{align}
which is e.g.~obtained by substituting Eq.~\eqref{eq:hittingTimeDist} into Eq.~\eqref{eq:bwFPEhittingTimeDist}. Using the definitions of the first passage probability and the mean first passage time one obtains the defining ODE for the first passage probability  \cite{gardiner1985handbook}
\begin{align}
\label{eq:ode_firstPassageProb}
0
 =
 f(\widehat{L})
 \frac{\partial}{\partial \widehat{L}} u(\widehat{L})
 +
 \frac{D(\widehat{L})}{2}
 \frac{\partial^2}{\partial \widehat{L}^2}
 u(\widehat{L})
  \quad \mathrm{with} \quad u(\widehat{L}_b)=1 \ , \ u(\widehat{L}_a)=0
  \ .
\end{align}
We here assumed $\widehat{L}_b$ to be reached first. The ODE for the mean first passage time  \cite{gardiner1985handbook}
\begin{align}
\label{eq:ode_meanFirstP}
-1
 =
 f(\widehat{L})
 \frac{\partial}{\partial \widehat{L}} v(\widehat{L})
 +
 \frac{D(\widehat{L})}{2}
 \frac{\partial^2}{\partial \widehat{L}^2}
 v(\widehat{L})
  \quad \mathrm{with} \quad v(\widehat{L}_a)=0 \ , \ v(\widehat{L}_b)=0
 \ .
\end{align}
follows analogously. \\

It is instructive to first solve these two expressions for a random walk before going into the general case. This will provide intuition for the meaning of the four class defining measures. The probability to reach $\widehat{L}_b$ before $\widehat{L}_a=0$ is
\begin{align}
  u(\widehat{L}) &= \frac{\widehat{L}}{\widehat{L}_b}
  \ .
  \intertext{The mean first exit time through either boundary for Brownian motion evaluates to}
  v(\widehat{L}) &= \frac{\widehat{L}}{D} (\widehat{L}_b -\widehat{L})
  \ .
  \intertext{As expected it is zero right at the boundaries and maximal right in the middle. Rewriting the mean first exit time to account for the contribution of each boundary separately}
&=
\frac{\widehat{L}}{\widehat{L}_b}
\cdot
\frac{(\widehat{L}_b - \widehat{L} )^2}{D}
+
\frac{\widehat{L}_b-\widehat{L}}{\widehat{L}_b}
\cdot
\frac{\widehat{L}^2}{D}
\\
&=
u(\widehat{L}) \; N(\widehat{L},\widehat{L}_b)
+
(1 - u(\widehat{L}) ) \; \Sigma(\widehat{L}_a,\widehat{L})
\end{align}
allows to explain the Feller boundary classification intuitively.
The first term represents the probability of first passage through $\widehat{L_b}$ times the average time it takes to get there starting from $\widehat{L}$. The average time it takes to reach the boundary is proportional to the distance to the boundary squared, i.e. the mean square displacement, as expected for a diffusive process. The second term analogously describes the probability of first passing through $\widehat{L}_a$ times the average duration the dynamics take to do so. 
Feller's boundary classification exploits that these two terms can also be rewritten to ask whether a single chosen boundary (say $\widehat{L}_a$) can be reached from the inside of the chosen interval on average in finite time $\Sigma(\widehat{L}_a,\widehat{L})$, or whether the boundary at $\widehat{L}_a$ can be exited in finite time $N(\widehat{L}_a,\widehat{L})$. The behavior of Brownian motion close to a boundary $\widehat{L}_a$ can thus be summarized as 'attainable' and 'regular'. \\

Let us now leave the example and return to the general case. As demonstrated for the most simple case of Brownian motion, the Feller boundary classification scheme is based on the interpretation of the formal solution of the two ODEs Eq.~\eqref{eq:ode_firstPassageProb} and Eq.~\eqref{eq:ode_meanFirstP}. While in principle easily to solve, it allows for a much deeper understanding of the result when relating the result to the more general theory of second order parabolic equations such as the backward FPE. This theory demonstrates how each 1d backward FPE equation can be mapped to Brownian motion, scaling first the position variables and applying a time change subsequently. The intuitive approach for pure Brownian motion, as stated above, can thus directly be transferred to general diffusion processes.

The scaling becomes most apparent when rewriting the propagator of the underlying backward FPE 
\begin{align}
\mathcal{L} p(\widehat{L})
 &=
 f(\widehat{L})
 \frac{\partial}{\partial \widehat{L}} p(\widehat{L})
 +
 \frac{D(\widehat{L})}{2}
 \frac{\partial^2}{\partial \widehat{L}^2}
 p(\widehat{L})
 \\
 &=
 \frac{1}{2m(\widehat{L})}
 \frac{d}{d\widehat{L}}
 \left(
 \frac{1}{s(\widehat{L})}
 \frac{d p(\widehat{L})}{d\widehat{L}}
 \right)
\end{align}
in terms of its scale density
\begin{align}
\label{scaledensity}
 s(\widehat{L})
 =
 \exp
 \left(
 -\int_{\widehat{L}_a}^{\widehat{L}} \frac{2 f(\xi)}{D(\xi)} d\xi
 \right)
\end{align}
and speed density
\begin{align}
\label{speeddensity}
 m(\widehat{L}) = \frac{1}{D(\widehat{L}) s(\widehat{L})}
 \ .
\end{align}
Introducing the scale function
\begin{align}
\label{scalefunction}
 S(\widehat{L}) = \int_{\widehat{L}_a}^{\widehat{L}} s(\eta)d\eta
\end{align}
and speed measure
\begin{align}
\label{speedmeasure}
 M(\widehat{L}) = \int_{\widehat{L}_a}^{\widehat{L}} m(\eta) d\eta
\end{align}
the propagator further simplifies to
\begin{align}
\label{eq:scaleSpeedBwdFPE}
 \mathcal{L} p(\widehat{L})
 =
 \frac{1}{2}
 \frac{d}{dM(\widehat{L})}
 \left(
 \frac{d p(\widehat{L})}{dS(\widehat{L})}
 \right)
 \ .
\end{align}
Solving this expression with the left hand side set to zero we obtain the first passage probability through $\widehat{L}_b$ as
\begin{align}
  u(\widehat{L}) 
  = 
 \frac{S(\widehat{L})-S(\widehat{L}_a)}{S(\widehat{L}_b)-S(\widehat{L}_a)}
 \ .
\end{align}
It is the very same expression as for Brownian motion with all coordinates scaled according to the scaling function.
Solving Eq.~\eqref{eq:scaleSpeedBwdFPE} again with the left hand side set to minus one  \cite{gardiner1985handbook}, we obtain the general solution of the mean first passage time
\begin{align}
v(\widehat{L}) &= 
 2 u(\widehat{L})
 \int_{\widehat{L}}^{\widehat{L}_b} (S(\widehat{L}_b)-S(y))m(y) dy
 \\
 &\quad
 +
 2 (1-u(\widehat{L}))
 \int_{\widehat{L}_a}^{\widehat{L}} ( S(y)-S(\widehat{L}_a) )m(y)dy
\end{align}
with its contribution from the upper (first term) and lower (second term) boundary.
Like for the Brownian motion example, the first term constitutes the probability to reach the upper boundary before the lower, times the average time it takes to get there and vice versa for the lower boundary.

Instead of using the full expression for $u(\widehat{L})$ and $v(\widehat{L})$, the Feller boundary classification scheme analyzes these two expression in meaningful chunks. To characterize whether the boundary can be reached at all it is sufficient to study the denominator of $u(\widehat{L})$ for the upper boundary and $1-u(\widehat{L})$ for the lower boundary respectively. If the denominator is finite, the nominator is finite as well as it is a monotonically increasing function. If it is infinite the probability of reaching the target boundary before the other is zero. We here only consider the case for for the lower boundary $\widehat{L}_a \to \widehat{L}_\mathrm{ts}$. The corresponding criterion is thus defined as whether the scale measure
\begin{align}
 S[\widehat{L}_\mathrm{ts},\widehat{L}_b]
 =
 S(\widehat{L}_b) - S(\widehat{L}_\mathrm{ts})
 =
 \lim_{\widehat{L}_a \to \widehat{L}_\mathrm{ts}} 
 S(\widehat{L}_a,\widehat{L}_b]
\end{align}
is finite or infinite.
Whether the lower boundary can be reached in finite time is defined by the integral
\begin{align}
 \Sigma[\widehat{L}_a,\widehat{L}_b]
 =
 \int_{\widehat{L}_a}^{\widehat{L}_b} S[\widehat{L}_a,\widehat{L}] dM(\widehat{L})
 =
 \int_{\widehat{L}_a}^{\widehat{L}_b} (S(\widehat{L})-S(\widehat{L}_a)) m(\widehat{L})d\widehat{L}
 \ .
\end{align}
Note that the integral to be finite for $\widehat{L}_a \to \widehat{L}_\mathrm{ts}$ requires $\Sigma[\widehat{L}_\mathrm{ts},\widehat{L}_b]$ to be finite.  
Whether the lower boundary can be exited in finite time is defined by the integral
\begin{align}
 N[\widehat{L}_a,\widehat{L}_b]
 =
 \int_{\widehat{L}_a}^{\widehat{L}_b} S[\widehat{L},\widehat{L}_b]dM(\widehat{L})
 =
 \int_{\widehat{L}_a}^{\widehat{L}_b} (S(\widehat{L}_b)-S(\widehat{L}))m(\widehat{L})d\widehat{L}
=\int_{\widehat{L}_a}^{\widehat{L}_b} M[\widehat{L}_a,\widehat{L}] dS(\widehat{L})
\ .
 \end{align}
In particular, expressing $N[\widehat{L}_a,\widehat{L}_b]$ in terms of the speed measure $M[\widehat{L}_a,\widehat{L}]$ shows that the integral can only be finite if $M[\widehat{L}_a,\widehat{L}]$ is finite. Evaluating $S[\widehat{L}_\mathrm{ts},\widehat{L}_b]$, $M[\widehat{L}_\mathrm{ts},\widehat{L}_b]$, $\Sigma[\widehat{L}_\mathrm{ts},\widehat{L}_b]$ and $N[\widehat{L}_\mathrm{ts},\widehat{L}_b]$
thus constitutes a sufficient set of expression to characterize the boundary behavior of 1d diffusion processes. A corresponding lookup table is provided in Table \ref{tab:SN_mean_var}.

\begin{table}[]
    \centering
\begin{tabular}{ |c|c|c| c|| c| }
\hline
\multicolumn{4}{|c||}{criteria } & \\
\hline
$S(\widehat{L}_a,\widehat{L}]$& $M(\widehat{L}_a,\widehat{L}]$ & $\Sigma(\widehat{L}_a,\widehat{L}]$ & $N(\widehat{L}_a,\widehat{L}]$ & Feller boundary class \rule[-2ex]{0pt}{6ex} \\
\hline
$< \infty$ & $< \infty$ & $< \infty$ &$< \infty$ & regular\\
$< \infty$ & $= \infty$ & $< \infty$ &$= \infty$ & exit \\
$< \infty$ &$= \infty$  & $= \infty$ &$= \infty$ & natural \\
$= \infty$  &$< \infty$ & $= \infty$ &$= \infty$ & natural\\
$= \infty$  &$= \infty$ & $= \infty$ &$= \infty$ & natural\\
$= \infty$  &$< \infty$ & $= \infty$ &$< \infty$ & entrance \\
\hline
\end{tabular}
    \caption{Decision criteria for Feller boundary classification. The class is determined by the asymptotic behavior of the criteria as the target state is approached $\widehat{L}_a \to \widehat{L}_\mathrm{ts}$.}
    \label{tab:SN_mean_var}
\end{table}


\subsection{Gamma functions and asymptotics}
Before the Feller boundary classification is applied to our TSA dynamics we need to mention a few expressions and asymptotic results for the upper incomplete gamma function 
\begin{align}
 \Gamma\left(s,x\right)
 =
 \int_x^\infty
 t^{s-1} e^{-t} dt
 \ ,
\end{align}
which often appears in the following analytic derivations. Confining this function to an interval one obtains
\begin{align}
 \int_{x_a}^{x_b} t^{s-1} e^{-t} dt
 =
 \Gamma\left(s,x_a\right)
 -
  \Gamma\left(s,x_b\right)
  =
  -
  \Gamma\left(s,x\right)
  \Big|_{x_a}^{x_b}
  \ .
\end{align}
All integrals of the following derivation are of the form
\begin{align}
  \int_{L_a}^{L_b} L^{n} e^{-cL^m} dL
  =
  \frac{c^{-\frac{n+1}{m}}}{m}  
  \int_{c L_a^m}^{c L_b^m} 
  t^{\frac{n+1}{m}-1}
  e^{-t} dt
  =
  -
   \frac{c^{-\frac{n+1}{m}} \Gamma\left(\frac{n+1}{m}, c L^m \right)}{m}
   \Big|_{L_a}^{L_b}
\end{align}
and can be evaluated by transformation to the upper incomplete Gamma function. As some of the integrals in the following derivation can not be evaluated analytically, and since we are only interested in the behavior at the boundary for the classification, we need to know the limit behavior of the upper incomplete gamma function. For the following derivations it is favourable to choose approximations which explicitly include a factor $e^{-z}$, which is why some approximations look slightly different than the standard textbook ones.

First, we consider $|cL^m| \to \infty$, and by using the known asymptotic expansion Eq.~(AS 6.5.32) \cite{abramowitz1964handbook} obtain
\begin{align}
\label{sup_GammaFktExpanInf}
\Gamma \left( \nu, z \right)
=
z^{\nu-1}  e^{-z}  
\left(1 + \frac{\nu-1}{z}   
+ \mathcal{O}(z^{-2})
\right)
\qquad \mathrm{for} \ \left| z \right| \to \infty
\ .
\end{align}
Secondly, we exploit the limit of $|cL^m| \to 0$ by using the series expansion
\begin{align}
\label{sup_GammaFktExpanZero}
\Gamma (\nu,z)
&=
\Gamma(\nu)
-
\Gamma(\nu) z^\nu e^{-z} \sum_{n=0}^{\infty} \frac{z^n}{\Gamma(\nu+n+1)}
\qquad 
\mathrm{for} \ |z| < \infty 
\notag \\
&=
\Gamma(\nu)
-
z^\nu e^{-z}
\left(
\frac{1}{\nu}
+
\frac{z}{\nu(\nu+1)} 
+
\mathcal{O}\left( z^2 \right)
\right)
%
\end{align}
obtained from Eq.(AS 6.5.3), Eq.(AS 6.5.4) and Eq.(AS 6.5.29) 
 \cite{abramowitz1964handbook}. The expression for $\Gamma (\nu)$ is not needed later, so we omit to state it here.\\
Thirdly, for $\nu = - n$ with $n \in \mathbb{N}_0$ the expansion of the upper incomplete gamma function is split to an expression for $n=0$, which, in accordance with (8.4.1-8.4.15) from the DLMF  \cite{DLMF}, reads
\begin{align}
\Gamma (0,z)&=\lim _{s\to 0}\left(\Gamma (s)-{\tfrac {1}{s}}-(\gamma (s,z)-{\tfrac {1}{s}})\right) \notag\\
&=-\gamma_E -\ln(z)-\sum _{k=1}^{\infty }{\frac {(-z)^{k}}{k\,(k!)}}
\ .
\end{align}
$\gamma_E$ is the Euler-Mascheroni constant, where we introduced the index $E$ to distinguish it from the parameter $\gamma$ used through out this text. $\gamma(s,z)$ is defined in Eq.(AS 6.5.3), Eq.(AS 6.5.4)
 \cite{abramowitz1964handbook}. We don't need its expression for the following derivations, so we don't explicitly state it here.

The expression for $n \in \mathbb{N}$ then reads
\begin{align}
\label{nGamma}
\Gamma (-n,z)={\frac {1}{n!}}\left({\frac {e^{-z}}{z^{n}}}\sum _{k=0}^{n-1}(-1)^{k}(n-k-1)!\,z^{k}+(-1)^{n}\Gamma (0,z)\right)
\ .
\end{align}

\subsection{Derivation of the boundary classification}
To classify the boundary behavior for one dimensional diffusion processes of the form
\begin{align}
 d\widehat{L} = f(\widehat{L}) dt +\sqrt{D(\widehat{L})} dW_t
\end{align}
close to a target state $\widehat{L}_\mathrm{ts} = 0$, and positive values for $\widehat{L}$,
we expand the advection and diffusion terms in their lowest order power law
\begin{align}
 f(\widehat{L}) &= - \gamma \widehat{L}^\alpha
 \\
  D(\widehat{L}) &= D \widehat{L}^\beta
\end{align}
with $\gamma \neq 0 $, $D>0$ and $\alpha,\beta \in \mathbb{R}$..

Under these assumptions, and using equations (\ref{scaledensity}) and (\ref{speeddensity}), the scale density 
\begin{align}
s(\widehat{L}) 
=
\begin{cases}
  \widehat{L}^{\frac{2 \gamma }{D}} 
  \qquad \qquad
&\ \alpha -\beta = -1
\\
e^{
\frac{ 2 \gamma  \widehat{L}^{1+\alpha-\beta}}{D(1+\alpha-\beta)} 
}
\qquad \qquad
&\ \alpha -\beta \neq -1 \ .
\end{cases}
\end{align}
and
speed density 
\begin{align}
m(\widehat{L})
=
\begin{cases}
 \frac{\widehat{L}^{-\alpha -\frac{2 \gamma }{D}-1}}{D}
   \qquad \qquad
&\ \alpha -\beta = -1
\\
\frac{\widehat{L}^{-\beta}}{D}
 e^{-\frac{2 \gamma  \widehat{L}^{1+\alpha -\beta}}{D (1+\alpha -\beta)}}
   \qquad \qquad
&\ \alpha -\beta \neq -1
\end{cases}
\end{align}
can be calculated. Integrating over the scale density the speed measure (\ref{speedmeasure}) reads
\begin{align} 
S(\widehat{L}_a,\widehat{L}_b)
=
&\int_{\widehat{L}_a}^{\widehat{L}_b} d\widehat{L}
\; s(\widehat{L})
\notag \\
&=
\begin{cases}
\int_{\widehat{L}_a}^{\widehat{L}_b} d\widehat{L} \;
\widehat{L}^{\frac{2 \gamma }{D}}
    \qquad \qquad
&\ \alpha -\beta = -1
\notag \\
\int_{\widehat{L}_a}^{\widehat{L}_b} d\widehat{L} \;
 e^{\frac{2 \gamma  \widehat{L}^{\alpha -\beta +1}}{D (\alpha -\beta +1)}} 
    \qquad \qquad
&\ \alpha -\beta \neq -1
\end{cases}
\notag \\
 &=
\begin{cases}
\frac{D }{2 \gamma +D}
{\widehat{L}}^{\frac{2 \gamma }{D}+1}
&\
\Big|_{\widehat{L}_a}^{\widehat{L}_b}
\qquad \mathrm{for} \quad
\alpha -\beta = -1
 \\ 
-
\frac{
    \left(-
    \frac{2 \gamma }{D(\alpha -\beta +1)}
   \right)^{-\frac{1}{\alpha -\beta +1}} 
        \Gamma \left(\frac{1}{\alpha -\beta
        +1},- \frac{2 \gamma {\widehat{L}}^{\alpha -\beta +1}}{D(\alpha -\beta +1)}\right)
   }{\alpha -\beta +1}
   &\
   \Big|_{\widehat{L}_a}^{\widehat{L}_b}
    \qquad \mathrm{for} \quad
\alpha -\beta \neq -1   
\ ,
\end{cases}
\intertext{where the full expression is analytically accessible. For interpretability we next expand $S(\widehat{L}_a,\widehat{L}_b)$. For $\alpha-\beta+1<0$ the argument 
$z= - \frac{2 \gamma {\widehat{L}}^{\alpha -\beta +1} }{D(\alpha -\beta +1)}$ of the upper incomplete gamma function $\Gamma\left(\nu,z\right)$ diverges for $\widehat{L} \to 0$ and $\gamma \neq 0$. We use the asymptotic expansion stated in Eq.~\eqref{sup_GammaFktExpanInf}. For $\alpha-\beta+1>0$, and $\gamma \neq 0$ one finds $z\to 0^-$. We use the expansion stated in Eq.~\eqref{sup_GammaFktExpanZero}. To lowest order we therefore find}
&=
\begin{cases}
e^{\frac{2 \gamma {\widehat{L}}^{\alpha -\beta +1}}{D(\alpha -\beta +1)}}
 \left(
\frac{D}{2 \gamma}
    {\widehat{L}}^{\beta -\alpha }
  +
  \mathcal{O}\left({\widehat{L}}^{-2 \alpha +2 \beta -1}\right)
  \;
  \right)  \,
  \Big|_{\widehat{L}_a}^{\widehat{L}_b}
   + \text{const.}
  \qquad &\mathrm{for} \quad \alpha-\beta < -1
 \\
\qquad \qquad \quad
\frac{D}{2 \gamma +D} {\widehat{L}}^{\frac{2 \gamma }{D}+1}
\qquad \qquad \qquad \quad \ \;
 \Big|_{\widehat{L}_a}^{\widehat{L}_b} 
  &\mathrm{for} \quad \alpha-\beta = -1
  \\
e^{\frac{2 \gamma {\widehat{L}}^{\alpha -\beta +1}}{D(\alpha -\beta +1)}}
\left(
\widehat{L} 
\qquad
\qquad
+\mathcal{O}\left(\widehat{L}^{\alpha -\beta +2}\right)
\quad
\right)
  \Big|_{\widehat{L}_a}^{\widehat{L}_b}
  + \text{const.}
  &\mathrm{for} \quad \alpha-\beta > -1
  \ .
\end{cases}
\end{align}
We only stated terms which are relevant for the asymptotic behavior and omitted writing other constant terms. All terms depending on $\widehat{L}_a$ in $S(\widehat{L}_a,\widehat{L}_b)$ thus approach zero for $\widehat{L}_a\to 0$ and we conclude
$|S(\widehat{L}_a,\widehat{L}_b)|<\infty$ $\forall$ $\alpha,\beta$ and $\gamma>0$. For $\gamma < 0$ we now obtain $|S(\widehat{L}_a,\widehat{L}_b)| \to \infty$ for $\alpha -\beta < -1$. On the boundary between the two regions for $\alpha -\beta = -1$ and $\gamma \leq -D/2$ we obtain $|S(\widehat{L}_a,\widehat{L}_b)| \to \infty$ and for $\gamma > -D/2$ we get  $|S(\widehat{L}_a,\widehat{L}_b)| < \infty$.

Next we evaluate the speed measure (\ref{speedmeasure})
\begin{align}
 M(\widehat{L}_a,\widehat{L}_b)
 &=
 \int_{\widehat{L}_a}^{\widehat{L}_b} d\widehat{L} \; m(\widehat{L})
 \notag \\
 &=
 \int_{\widehat{L}_a}^{\widehat{L}_b} d\widehat{L} \;
 \begin{cases}
\frac{\widehat{L}^{-\alpha -\frac{2 \gamma }{D}-1}}{D}
 \qquad \qquad \
 \mathrm{for} \quad \alpha-\beta = -1
 \notag \\ 
  \frac{\widehat{L}^{-\beta}}{D}
 e^{-\frac{2 \gamma  \widehat{L}^{1+\alpha -\beta}}{D (1+\alpha -\beta)}}
 \qquad \mathrm{for} \quad \alpha-\beta \neq -1
 \end{cases}
 \notag \\
 &=
 \begin{cases}
 \begin{cases}
 \frac{1}{D}\log \widehat{L}
 \ \
  \Big|_{\widehat{L}_a}^{\widehat{L}_b}
&\quad  \alpha = -\frac{2\gamma}{D}
 \\
 -
 \frac{1}{2 \gamma +\alpha  D}
 {\widehat{L}}^{-\alpha -\frac{2 \gamma }{D}} 
 \
 \Big|_{\widehat{L}_a}^{\widehat{L}_b}
 &\quad  \mathrm{else}
 \end{cases}
 &\qquad \mathrm{for} \quad \alpha-\beta = -1
 \\
 -
 \frac{ \left(\frac{2\gamma }{D(\alpha -\beta 
   +1)}\right)^{\frac{\alpha }{\alpha -\beta +1}} 
   \Gamma \left(1 - \frac{\alpha }{\alpha -\beta+1},\frac{2 \gamma {\widehat{L}}^{\alpha -\beta +1} }{D(\alpha -\beta  +1)}\right)}{2 \gamma }
   \
    \Big|_{\widehat{L}_a}^{\widehat{L}_b}
   &\qquad \mathrm{for} \quad \alpha-\beta \neq -1 \ ,
\end{cases}
\intertext{which can be done exactly. For interpretability we expand $M(\widehat{L}_a,\widehat{L}_b)$. For $\alpha-\beta+1<0$ the argument 
$z=\frac{2 \gamma {\widehat{L}}^{\alpha -\beta +1}}{D(\alpha -\beta +1)}$ of the upper incomplete gamma function $\Gamma\left(\nu,z\right)$ diverges for $\widehat{L} \to 0$ and $\gamma \neq 0$. We use the asymptotic expansion stated in Eq.~\eqref{sup_GammaFktExpanInf}. For $\alpha-\beta+1>0$ and $\gamma\neq 0$ one finds $z\to 0^+$. We use the expansion stated in Eq.~\eqref{sup_GammaFktExpanZero}. The first argument of the gamma function $\nu = 1- \frac{\alpha}{\alpha-\beta+1}$ assumes negative whole number values for $\beta = 1+\frac{n-1}{n}\alpha, n \in \mathbb{N}$. For those cases we use the expansion stated in \eqref{nGamma}. To lowest order we find}
\label{eq:Mapprox}
 &=
\begin{cases}
 e^{-\frac{2 \gamma {\widehat{L}}^{\alpha -\beta +1}}{D(\alpha -\beta +1)}}
\left(
 -
\frac{1}{2 \gamma}
    {\widehat{L}}^{-\alpha }
    \quad \
  +
  \mathcal{O}\left({\widehat{L}}^{-2 \alpha + \beta -1}\right)
  \;
  \right)  
  \Big|_{\widehat{L}_a}^{\widehat{L}_b}
  &\qquad \mathrm{for} \quad \alpha-\beta < -1
\\
\begin{cases}
\frac{1}{D} \log \widehat{L}
\ \
\Big|_{\widehat{L}_a}^{\widehat{L}_b}
&\quad  \alpha = -\frac{2\gamma}{D}
\\
 -
 \frac{{\widehat{L}}^{-\alpha -\frac{2 \gamma }{D}} }{2 \gamma +\alpha  D}
 \Big|_{\widehat{L}_a}^{\widehat{L}_b}
 &\quad  \mathrm{else}
\end{cases}
 &\qquad \mathrm{for} \quad \alpha-\beta = -1
\\
\begin{cases}
- e^{-\frac{2 \gamma {\widehat{L}}^{\alpha -\beta +1}}{D(\alpha -\beta +1)}}\frac{1}{2\gamma n}  \widehat{L}^{-\alpha} 
+\mathcal{O}\left(\widehat{L}^{\frac{1-n}{n}\alpha}\right)
\ \
  \Big|_{\widehat{L}_a}^{\widehat{L}_b}
&\quad \beta=1+\frac{n-1}{n}\alpha 
\\
 e^{-\frac{2 \gamma {\widehat{L}}^{\alpha -\beta +1}}{D(\alpha -\beta +1)}}
\left(
\frac{1}{D(1-\beta)}
\widehat{L}^{1-\beta} 
+\mathcal{O}\left(\widehat{L}^{\alpha -2\beta +2}\right)
\right)
\ \
  \Big|_{\widehat{L}_a}^{\widehat{L}_b} 
    +\text{const.}
&\quad \mathrm{else}
\end{cases}
&\qquad \mathrm{for} \quad \alpha-\beta > -1
  \ .
\end{cases}
\end{align}
$n$ is always an integer not including zero, $n \in \mathbb{N}$. Constant values that do not contribute to the asymptotic behavior of the expressions are omitted. We first evaluate the asymptotic behavior $\widehat{L}_a \to 0$ in the case $\gamma >0$. For $\alpha-\beta<-1$ we find $|M(\widehat{L}_a,\widehat{L}_b)| \to \infty$. For $\alpha-\beta>-1$  we find $|M(\widehat{L}_a,\widehat{L}_b)| < \infty$ for $\beta<1$ and $|M(\widehat{L}_a,\widehat{L}_b)| \to \infty$ for $\beta \geq 1$. On the boundary for $\alpha-\beta=-1$ we find $|M(\widehat{L}_a,\widehat{L}_b)| \to \infty$ for $\alpha > -2\gamma/D$ and $|M(\widehat{L}_a,\widehat{L}_b)| < \infty$ else. For $\gamma < 0$, the situation stay the same for $\alpha-\beta>-1$. For $\alpha-\beta<-1$ we now find $|M(\widehat{L}_a,\widehat{L}_b)| < \infty$ and on the connecting line $\alpha-\beta=-1$ we find $|M(\widehat{L}_a,\widehat{L}_b)| \to \infty$ for $\alpha > -2\gamma/D$ and $|M(\widehat{L}_a,\widehat{L}_b)| < \infty$ else. Note that $\gamma$ now is negative, therefore the behavior on the connecting line is mirrored in comparison the case of positive $\gamma$.

Next we evaluate whether the lower boundary can on average be reached in finite time by computing $\Sigma$ and $N$. We calculate
\begin{align}
 \Sigma(\widehat{L}_a,\widehat{L}_b)
 &=
  \int_{\widehat{L}_a}^{\widehat{L}_b} d\widehat{L} \; S(\widehat{L}_a,\widehat{L}) m(\widehat{L})
  \ .
\intertext{This form of the expression reveals that the integral can only be evaluated for $|S(\widehat{L}_a,\widehat{L})|>0$, which is the case $\forall$ $\alpha,\beta$ and $\gamma \neq 0$. We rewrite the expression as}
&=
 \int_{\widehat{L}_a}^{\widehat{L}_b} d\widehat{L} \; M(\widehat{L},\widehat{L}_b) s(\widehat{L})
\ .
\intertext{
We observe that $\int_{\widehat{L}_a}^{\widehat{L}_b} d\widehat{L} \; M(\widehat{L}_b) s(\widehat{L})= M(\widehat{L}_b) S(\widehat{L}_a,\widehat{L}_b)$ and thus its behavior for $\widehat{L}\to \widehat{L}_a$ is purely defined by $S(\widehat{L}_a,\widehat{L}_b)$ which is always finite. We therefore use the approximation for $M(\widehat{L},\widehat{L}_b)$ stated in Eq.\eqref{eq:Mapprox} and  find
}
 &=
  M(\widehat{L}_b)S(\widehat{L}_a,\widehat{L}_b) - \int_{\widehat{L}_a}^{\widehat{L}_b} d\widehat{L} \; M(\widehat{L}) s(\widehat{L})
\intertext{We here see why we chose approximations of the upper incomplete gamma function $\Gamma(\nu, z)$ that include a factor $e^{-z}$, because those factors cancel when $M(\widehat{L})$ and $s(\widehat{L})$ are taken together. This yields}
&=
M(\widehat{L}_b)S(\widehat{L}_a,\widehat{L}_b) -
\int_{\widehat{L}_a}^{\widehat{L}_b} d\widehat{L} \;
\begin{cases}
 \left(
 -
\frac{1}{2 \gamma}
    {\widehat{L}}^{-\alpha }
    \quad \
  +
  \mathcal{O}\left({\widehat{L}}^{-2 \alpha + \beta -1}\right)
  \;
  \right)  
  &\qquad \mathrm{for} \quad \alpha-\beta < -1
\\
\begin{cases}
\frac{1}{D} \widehat{L}^{-\alpha} \log \widehat{L}
&\quad  \alpha = -\frac{2\gamma}{D}
\\
 -
 \frac{1}{2 \gamma +\alpha  D} {\widehat{L}}^{-\alpha}
 &\quad  \mathrm{else}
\end{cases}
 &\qquad \mathrm{for} \quad \alpha-\beta = -1
\\
\begin{cases}
-\frac{1}{2\gamma n}  \widehat{L}^{-\alpha} 
+\mathcal{O}\left(\widehat{L}^{\frac{1-n}{n}\alpha}\right)
\ \
&\quad \beta = 1 + \frac{n-1}{n}\alpha
\\
\left(
\frac{1}{D(1-\beta)}
\widehat{L}^{1-\beta} 
+\mathcal{O}\left(\widehat{L}^{\alpha -2\beta +2}\right)
\right)
&\quad  \mathrm{else}
\end{cases}
 \ \ &\qquad \mathrm{for} \quad \alpha-\beta > -1
\end{cases}
\notag \\
&=
M(\widehat{L}_b)S(\widehat{L}_a,\widehat{L}_b) 
-
\begin{cases}
 \begin{cases}
 \left(
 -\frac{1}{2\gamma} \log \widehat{L} 
   +
  \mathcal{O}\left({\widehat{L}}^{-2 \alpha + \beta}\right)
  \;
  \right)  
  \Big|_{\widehat{L}_a}^{\widehat{L}_b} 
 \ \
&\quad \alpha = 1
 \\
 \left(
 -
\frac{1}{2 \gamma (1-\alpha)}
    {\widehat{L}}^{1-\alpha }
    \quad \
  +
  \mathcal{O}\left({\widehat{L}}^{-2 \alpha + \beta}\right)
  \;
  \right)  
  \Big|_{\widehat{L}_a}^{\widehat{L}_b} 
 \ \
&\quad \mathrm{else} 
 \end{cases}
  &\qquad \mathrm{for} \quad \alpha-\beta < -1
\\
\begin{cases}
 -\frac{\widehat{L}^{1-\alpha }}{(1-\alpha ) (2 \gamma +D)}
 + \frac{\widehat{L}^{1-\alpha } \log (\widehat{L})}{(1-\alpha) D}
 \ \
&\quad \alpha = -\frac{2\gamma}{D}
\\
\begin{cases}
 -
 \frac{1}{(2 \gamma + D)} \log \widehat{L}
\ \  \Big|_{\widehat{L}_a}^{\widehat{L}_b}
\ \
&\quad \alpha = 1
 \\
 -
 \frac{1}{(2 \gamma +\alpha  D)(1-\alpha)} {\widehat{L}}^{1-\alpha}
\ \  \Big|_{\widehat{L}_a}^{\widehat{L}_b}
\ \
&\quad \mathrm{else}
\end{cases}
\end{cases}
%
%
&\qquad \mathrm{for} \quad \alpha-\beta = -1
\\
\begin{cases}
-\frac{1}{2 \gamma n} \log \widehat{L}
+
\mathcal{O} \left( \widehat{L}^{\frac{1}{n}} \right) 
\ \
  \Big|_{\widehat{L}_a}^{\widehat{L}_b} \qquad \qquad \qquad \, \, \, \alpha = 1
&\quad \beta = 1 + \frac{n-1}{n}\alpha
\\
-\frac{1}{2 \gamma n(1-\alpha)} \widehat{L}^{1-\alpha} 
+ 
\mathcal{O}\left(\widehat{L}^{\frac{1-n}{n}\alpha + 1} \right)
\ \
  \Big|_{\widehat{L}_a}^{\widehat{L}_b} \qquad \text{else}
&\quad  \beta = 1 + \frac{n-1}{n}\alpha
\\
\left(
\frac{1}{D(1-\beta)(2-\beta)}
\widehat{L}^{2-\beta} 
+\mathcal{O}\left(\widehat{L}^{\alpha -2\beta +3}\right)
\right)
\Big|_{\widehat{L}_a}^{\widehat{L}_b} 
&\quad  \mathrm{else}
\end{cases}
 &\qquad \mathrm{for} \quad \alpha-\beta > -1
\end{cases}
\end{align}
For the cases where we distinguish between different $\beta$ cases we always have $n \in \mathbb{N}$. We are allowed to evaluate the asymptotic behavior under the integral because all expressions under the integral are either divergent or dominated in the sense of Lebesgue's dominated convergence theorem  \cite{bookIntegration}. The same is true for the evaluation of $N(\widehat{L}_a,\widehat{L}_b)$ below For $\widehat{L}_a \to 0$ we again first evaluate the case $\gamma >0$. We find $|\Sigma(\widehat{L}_a,\widehat{L}_b)| < \infty$ for all regions except for a top right corner region which is given by $\beta >2$ and $\alpha>1$. In those cases $|\Sigma(\widehat{L}_a,\widehat{L}_b)|$ diverges. For negative $\gamma <0$ the expressions show exactly the same asymptotic behavior as for $\gamma >0$, however because $|S(\widehat{L}_a,\widehat{L}_b)|$ diverges for $\alpha - \beta < -1$, $|\Sigma(\widehat{L}_a,\widehat{L}_b)|$ diverges as well.

In the final step we evaluate whether the lower boundary can on average be exited in finite time. We evaluate
\begin{align}
 N(\widehat{L}_a,\widehat{L}_b)
 &=
 \int_{\widehat{L}_a}^{\widehat{L}_b} d\widehat{L} \; M(\widehat{L}_a,\widehat{L}) s(\widehat{L})
\ .
\intertext{
With $M(\widehat{L}_a,\widehat{L})$ explicitly depending on the lower boundary the integral ins only finite where $M(\widehat{L}_a,\widehat{L})$ is finite. This implies that we only need to check whether $N(\widehat{L}_a,\widehat{L}_b)$ is finite for $\alpha-\beta >-1$ and $\beta \le 1$. For completeness we nevertheless evaluate all terms starting from
}
&=
 \int_{\widehat{L}_a}^{\widehat{L}_b} d\widehat{L} \; S(\widehat{L},\widehat{L}_b) m(\widehat{L})
 =
 S(\widehat{L}_b) M(\widehat{L}_a,\widehat{L}_b) - 
 \int_{\widehat{L}_a}^{\widehat{L}_b} d\widehat{L} \; S(\widehat{L}) m(\widehat{L})
 \notag \\
&=
 S(\widehat{L}_b) M(\widehat{L}_a,\widehat{L}_b) 
 - 
\int_{\widehat{L}_a}^{\widehat{L}_b} d\widehat{L} \;
\begin{cases}
 \left(
\frac{1}{2 \gamma}
    {\widehat{L}}^{ -\alpha }
  +
  \mathcal{O}\left({\widehat{L}}^{-2 \alpha + \beta -1}\right)
  \;
  \right)  
  \qquad \mathrm{for} \quad \alpha-\beta < -1
\notag \\
\frac{1}{2 \gamma +D} \widehat{L}^{-\alpha}
\qquad \qquad \qquad \quad \ \;
  \qquad \mathrm{for} \quad \alpha-\beta = -1
\notag \\
\left(
\frac{1}{D}
\widehat{L}^{1-\beta} 
\qquad
\qquad
+\mathcal{O}\left(\widehat{L}^{\alpha -2\beta +2}\right)
\right)
\qquad \mathrm{for} \quad \alpha-\beta > -1
\end{cases}
 \notag \\
&=
 S(\widehat{L}_b) M(\widehat{L}_a,\widehat{L}_b) 
 - 
\begin{cases}
\begin{cases}
 \left(
\frac{1}{2 \gamma}
    \log \widehat{L}
  +
  \mathcal{O}\left({\widehat{L}}^{-2 \alpha + \beta}\right)
  \;
  \right)  
  \Big|_{\widehat{L}_a}^{\widehat{L}_b}
&\quad \alpha = 1
\\
 \left(
\frac{1}{2 \gamma (1-\alpha)}
    {\widehat{L}}^{1 -\alpha}
  +
  \mathcal{O}\left({\widehat{L}}^{-2 \alpha + \beta}\right)
  \;
  \right)  
  \Big|_{\widehat{L}_a}^{\widehat{L}_b}
&\quad \mathrm{else}  
\end{cases}
  &\qquad \mathrm{for} \quad \alpha-\beta < -1
\notag \\
\begin{cases}
\frac{1}{2 \gamma +D}
    \log \widehat{L}
    \ \;
  \Big|_{\widehat{L}_a}^{\widehat{L}_b}
&\quad \alpha = 1
\\
\frac{1}{(2 \gamma +D)(1 -\alpha)} \widehat{L}^{1 -\alpha}
\ \;
  \Big|_{\widehat{L}_a}^{\widehat{L}_b}
&\quad \mathrm{else}   
\end{cases}
  &\qquad \mathrm{for} \quad \alpha-\beta = -1
\notag \\
\begin{cases}
\frac{1}{D}
    \log \widehat{L}
 +\mathcal{O}\left(\widehat{L}^{\alpha -2\beta +3}\right)   
    \ \;
  \Big|_{\widehat{L}_a}^{\widehat{L}_b}
&\quad \beta = 2
\\
\left(
\frac{1}{D(2-\beta)}
\widehat{L}^{2-\beta} 
+\mathcal{O}\left(\widehat{L}^{\alpha -2\beta +3}\right)
\right)
  \Big|_{\widehat{L}_a}^{\widehat{L}_b} 
 &\quad \mathrm{else} 
\end{cases}
&\qquad \mathrm{for} \quad \alpha-\beta > -1
\end{cases}
 \end{align}
 For $\widehat{L}_a \to 0$ we first evaluate the case $\gamma >0$. We find that the second term in the expression for $|N(\widehat{L}_a,\widehat{L}_b)|$ is $<\infty$ for all regions except for a top right corner region which is given by $\beta >2$ and $\alpha>1$. Therefore, also the full expression for $|N(\widehat{L}_a,\widehat{L}_b)|$ behaves the same except for $\alpha -\beta < -1$ where $|M(\widehat{L}_a,\widehat{L}_b)|$ diverges and therefore $|N(\widehat{L}_a,\widehat{L}_b)|$ diverges as well. For negative $\gamma <0$ the expressions show exactly the same asymptotic behavior as for $\gamma >0$ except $|M(\widehat{L}_a,\widehat{L}_b)| < \infty$ for $\alpha -\beta < -1$ and thus $|N(\widehat{L}_a,\widehat{L}_b)| < \infty$ as well in this region.

One final question considers whether the boundary behavior changes when we start far away from the boundary. The in between region is covered as we integrate over densities (see paper where this idea is made explicit). The limit $\widehat{L} \to \infty$ is equivalent to switching the results for $\alpha-\beta > -1$ and $< -1$. As we stay inside of the interval $(0,\infty)$ none of the terms diverges and we have everywhere regular boundary conditions. This implies the boundary behavior we discussed is universal.



\bibliography{literature_SI}